\documentclass[%
aip,
jmp,%
amsmath,amssymb,
reprint,%
nofootinbib
]{revtex4-2}

\usepackage{graphicx}
\usepackage{dcolumn}
\usepackage{bm}

\usepackage{float} 
\usepackage{ulem}

\usepackage{parskip}
\usepackage{xcolor}
\newcommand{\g}[1]{\textcolor{black}{#1}}

\newcommand{\f}[1]{\textcolor{black}{#1}} 
\newcommand{\fs}[1]{\textcolor{black}{#1}} 

\begin{document}
	
	\title{Correlated collisions and history filtering: unraveling and reproducing the statistics of coalescing particles in turbulence from the ghost-particle framework}
	
	\author{Fanxi Gong}
	
	\affiliation{School of Atmospheric Sciences, Sun Yat-Sen University, Zhuhai, China.}

	\affiliation{Guangdong Province Key Laboratory for Climate Change and Natural Disaster Studies, Sun Yat-Sen University, Zhuhai, China.}
	
	\author{Ewe-Wei Saw}
	
	\email[Corresponding author: ]{ewsaw3@gmail.com}
	
	\altaffiliation{Also at: Ministry of Education Key Laboratory of Tropical Atmosphere-Ocean System, Zhuhai, China.}
		
	\affiliation{School of Atmospheric Sciences, Sun Yat-Sen University, Zhuhai, China.}

	\affiliation{Guangdong Province Key Laboratory for Climate Change and Natural Disaster Studies, Sun Yat-Sen University, Zhuhai, China.}

	\date{\today}
	\begin{abstract}
	 
	This is the first in a series of papers aimed at understanding and predicting the statistics of coalescing particles in turbulent flow, and their relation to the physics of the simpler and better-understood system of collisionless ghost particles in turbulence. We perform three distinct families of Direct Numerical Simulations (DNS) under identical flow conditions. The first family involves ghost particles that do not mutually interact; the second contains particles (monomers) that coalesce upon collision, with lost monomers replenished by injection of new particles at random positions to maintain a constant monomer density (CR); the third family involves particles with the same collision--coalescence kinetics but without replenishment of lost monomers (CN). All analyses are monodisperse---only the statistics and physics of the monomers are considered. Across the range of Stokes number studied ($St =0.01\text{--}3.0$), we find that the collision kernel ($K$) and the values of radial distribution function (RDF, $g(r)$) near particle contact ($r \approx d$, where $d$ is the particle diameter) of the ghost-particle system are higher than those of the coalescing systems. We show that a velocity-filtered version of the RDF $g_{\mathrm{G}}^{\scriptscriptstyle (-)}(r)$ is a reasonable proxy for estimating the CR system's RDF value at contact $g_{\mathrm{CR}}(d)$. We systematically report on the residual discrepancy between $g_{\mathrm{G}}^{\scriptscriptstyle (-)}(d)$ and $g_{\text{CR}}(d)$ as well as between the corresponding kernels $K_{\text{G}}$ and $K_{\text{CR}}$. We show that the discrepancy is due to correlation among successive collisions within the ghost-particle system---by showing that a history-filtered version of the ghost-particle collision kernel, which is void of such correlations, reproduces the corresponding kernel of the coalescing systems. We introduce a collision-correlation time, $\tau_{\text{cc}}$, to quantify how long current collisions influence future collision events and show that it is finite and has a narrow range across the range of $St$ studied. We show that when particles with past collisions within a period comparable to $\tau_{\text{cc}}$ are filtered out in the ghost-particle system, the resulting history-filtered RDF $\tilde{g}_{\text{G}}(r;\,\tau \approx  {\tau_{\text{cc}}})$ reproduces the RDF of the coalescing systems ($g_{\text{CR}}(r)$ and $g_{\text{CN}}(r)$). Finally, we show that among all history-bearing collisions in the ghost-particle system, the fraction of repeated collisions involving identical particles is an exponentially decaying function of $St$. This work unifies the coalescing and ghost particle systems under a single framework and affords an elegant interpretation of the former as a history filtered equivalent of the latter. 
	\end{abstract}
	
	\maketitle
	
 	\section{INTRODUCTION} 
	 Dynamical interactions between particle and turbulent flow, important for processes such as transport, mixing, and collisions of particles in such a flow, are ubiquitous. They can be seen in problems such as cloud droplet growth in the atmosphere\cite{Shaw2003,Grabowski2013} and spray drying or powder separation in industrial processes\cite{Crowe2005}. At the same time, the importance of these and similar particle-laden flows to both  natural and technological systems has been well established\cite{Balachandar2010,Kuerten2016}. At the heart of particle-laden turbulence lies the inter-particle collision rate ($\Gamma$), which ultimately controls outcomes such as coalescence, growth. 
	 From collision rate, one could define a descriptor known as the collision kernel ($K$), that encapsulates how the rate intrinsically depends on turbulent and particle-microphysical mechanisms\cite{Saffman1956,Sundaram1997,Wang2000}. Consequently, an accurate quantification of $K$ is indispensable for predictive modeling and for advancing the fundamental understanding of particle-laden systems\cite{Pan2010,Rosa2013}.
	 
	 Early theoretical efforts by Saffman and Turner\cite{Saffman1956} established that small-scale fluid velocity gradients induce relative motion between fully entrained particles, thereby driving their collisions. However, this model neglects particle inertia. Subsequent studies demonstrated that inertial effects cause particles to deviate from fluid streamlines, leading to preferential concentration in strain-dominant regions\cite{Maxey1987, Squires1991, Eaton1994, Toschi2009}. Such clustering increases the local particle density and thus the collision rate, and may be properly quantified by the radial distribution function (RDF) \cite{Sundaram1997, Wang2000, Monchaux2012, Saw2012}. In addition, the sling or velocity-caustic effect—where multivalued particle velocity fields form in phase space—enhances relative velocities between particles and collision rate\cite{Falkovich2002, Wilkinson2006, Saw2014, Pumir2016}. Within this framework, the collision kernel ($K$) is formally defined as the inward flux of particle pair crossing the collision surface at contact separation, such that the product of $K$ and particle density squared gives the collision rate per unit volume\cite{Saffman1956,Sundaram1997,Wang2000, Pruppacher1998}. For a monodisperse particulate system, many authors relate the collision kernel related to RDF and mean relative velocity via: $K = 4\pi d^2 g(d) \langle w_r(d) \rangle$, where $d$ is the particle diameter, $g(d)$ is the radial distribution function evaluated at contact, and $\langle w_r(d) \rangle$ denotes the mean of the radial component of relative velocity (mean radial velocity, MRV) of particle pairs at contact. 
	 
	 Direct numerical simulation (DNS) has played a central role in quantifying these statistics and in elucidating the turbulent mechanisms underlying clustering, caustic formation, and enhanced relative motion\cite{Reade2000,Ayala2008, Bec2010}. Within this context, many DNS are carried out with ghost particles that do not interact during collisions and would continue with their trajectories as if other particles are absent. These ghost particle simulations provide an idealized baseline, allowing comparison and quantitative validation of fundamental statistical theories\cite{Bragg2014, Gustavsson2016}. This approach further enables the elucidation of different mechanisms that enhance particle collision rate such as clustering and caustics\cite{Falkovich2002, Gustavsson2016}. However, a key shortcoming of this ghost particle approach lies in the fact that the ghost particles are unaffected by any collisions and thus every time their surface touches that of another particle, the event is counted as a new collision, thereby allowing each particle to collide multiple times with other particles.  As a result, it is known to systematically overestimate actual collision rates\cite{Gustavsson2008}, with pronounced discrepancies reported especially at low Stokes number conditions\cite{Vokuhle2011, Vokuhle2013}. This overprediction decreases as $St$ increases, reflecting the transition from clustering dominated to caustic dominated collision regimes\cite{Vokuhle2014}.

	 Beyond this inherent limitation, a deeper physical insight emerges when actual collision-induced particle interaction is taken into account. One example of significant interest is the coalescence (coagulation) of small droplets (particles) in turbulent flow. Saw and Meng\cite{Saw2022} showed that collision--coalescence sharply reduces the RDF at the particle-contact scale while simultaneously enhancing the magnitude of MRV. Consequently, while ghost particle simulations remain a valuable theoretical baseline, new simulations that resolve collision--coalescence dynamics highlight a persistent gap\cite{Saw2022, Meng2023, Meng2026}.

	 Despite its shortcomings, the ghost particle (GP) system has some advantages from the perspective of theoretical works owing its simplicity. Over the years, there has been significant theoretical understanding pertaining to it\cite{Saffman1956, Chun2005, Zaichik2005}.  Therefore, one major motivation of the current work is to explore the possibility of a theoretical bridge that connects available theories on the GP system to the more realistic case of system with coalescing particles in turbulent flow. As an initial step toward this objective, this work focuses on a systematic comparison of the two systems and the establishment of a foundational theoretical framework.

	To that end, our work explicitly considers both ghost particle and collision--coalescence simulations containing monodisperse particles. 
	We conduct a systematic comparison between the two types of system (simulation) over a range of Stokes numbers. Since coalescence would inadvertently reduces the number of monomers that we initially injected into the system, to isolate the role of time-variation of particle population size, we examine two sub-types of coalescing systems, one with particle replenishment (coalescing-replenished, CR) and one without (coalescing--non-replenished, CN). The comparison focuses on key statistical measures of particle-turbulence interactions—the collision kernel and the radial distribution function—while the mean radial relative velocity is analyzed as a factor linking these quantities. Particular emphasis is placed on the role of correlated successive collisions  pertaining to the GP system and its effect on key statistics. We emphasize that successive collisions are not restricted to repeated encounters between the same pair of particles.

	The remainder of this paper is organized as follows. Section~\ref{sec:methods} describes the numerical methods, including direct numerical simulation of Navier--Stokes turbulence and particle dynamics. Section~\ref{sec:postprocessing} outlines the simulation setup and statistical analysis. Section~\ref{sec:prelude} presents a heuristic discussion of an earlier model. Section~\ref{sec:res}  presents results and discussion. Finally, Section~\ref{sec:con} summarizes our conclusions.
	
	\section{NUMERICAL METHODS}\label{sec:methods}
	The following assumptions are adopted throughout this study. The fluid phase is statistically stationary and isotropic, sustained by low-wavenumber stochastic forcing in a triply periodic cubic domain.
	Only monodisperse particles are considered. In the coalescing simulations, colliding particles are removed immediately upon collision so that the particle population remains monodisperse throughout the simulation.
	Particles are treated within the point-particle approximation, with finite-size effects considered only during collision detection.
	The particle volume fraction varies from $3.5\times10^{-6}$ to $3.8\times10^{-5}$ and the mass loading does not exceed $\mathcal{O}(10^{-2})$, justifying the one-way coupling approximation \cite{elghobashi1994predicting}.
	Microscale hydrodynamic interactions (e.g., lubrication forces) and gravitational settling are neglected.
	The effects of these additional physical mechanisms are beyond the scope of the present study and will be examined in future work.
	Coalescence is assumed to occur instantaneously when the particle separation $r$ equals the particle diameter $d$, such that the active radial distribution function is zero for all separations $r \le d$.
	\subsection{Direct Numerical Simulation of Navier--Stokes Turbulence}
	
	The fluid phase was investigated through direct numerical simulation of particle-laden isotropic turbulence, which was performed using a pseudo-spectral method \cite{Canuto1988,Mortensen2016} to solve the incompressible Navier--Stokes equations:
	\begin{align}
		\frac{\partial \mathbf{u}}{\partial t} + (\mathbf{u} \cdot \nabla) \mathbf{u} 
		&= -\frac{1}{\rho} \nabla p + \nu \nabla^2 \mathbf{u} + \mathbf{f}(\mathbf{x}, t), \label{eq:ns_momentum} \\
		\nabla \cdot \mathbf{u} &= 0, \label{eq:ns_continuity}
	\end{align}
	where $\mathbf{u}$ is the velocity, $p$ is the pressure, $\rho$ is the density, and $\nu$ is the kinematic viscosity. 
	An external stochastic forcing term, $\mathbf{f}(\mathbf{x}, t)$, was incorporated into the momentum equation, exclusively activating at low wavenumbers to continuously inject kinetic energy \cite{Eswaran1988}. The computation was conducted within a triply periodic cube domain of side length $2\pi$, discretized over $256^3$ grid points. Aliasing errors arising from the nonlinear convection term were removed using the 2/3 dealiasing rule \cite{Rogallo1981}. The spatial resolution satisfies $k_{\max}\eta \approx 1.2$, where $k_{\max}=\sqrt{2}N_{\text{grid}}/3$ is the maximum resolved wavenumber magnitude. 
    This exceeds the commonly accepted condition $k_{\max}\eta > 1$ for adequately resolving the dissipative scales\cite{Eswaran1988}.  
    Furthermore, Meng and Saw (2023)\cite{Meng2023} performed a higher-resolution comparison using a grid resolution four times higher under the same DNS configuration, and reported that unresolved sub-Kolmogorov intermittency introduces only slight changes in the RDF without affecting its qualitative trends. The simulation methodology used here is closely similar to the one in Meng and Saw (2023)\cite{Meng2023}.
    The statistically stationary flow corresponds to a Taylor-scale Reynolds number of $Re_\lambda = 124$.
	Time advancement uses a second-order Runge–Kutta scheme with a constant time step $\Delta t \approx \tau_\eta / 100$, ensuring a maximum Courant number of approximately 0.069.
	
	\begin{table}[h]
		\caption{Simulation parameters for the fluid phase. All dimensional parameters are in arbitrary units. $N_{\text{grid}}$ is the simulation grid size, $\nu$ is the kinematic viscosity of turbulence, $\epsilon$ is the dissipation rate of turbulent flow, $u'$ is the root mean square velocity of turbulent flow, $\lambda$ is the Taylor length scale, $\eta$ and $\tau_\eta$ are the Kolmogorov length and timescale, respectively, $L = \frac{3\pi}{4} \int [E(k)/k] \, dk \big/ \int E(k)\, dk$ and $\tau_L = L/u'$ are the (longitudinal) integral length scale and large-eddy turnover timescale 
        respectively ($E(k)$ is the turbulent energy spectrum)\cite{pope2001turbulent}, and $Re_\lambda$ is the Taylor scaled Reynolds number.} 
		\label{tab:fluid_params}
		\centering
		\begin{ruledtabular}
			\renewcommand{\arraystretch}{1.0}
			\begin{tabular}{cccccccccc}
				$N_{\text{grid}}$ & $\nu$ & $\epsilon$ & $u'$ & $\lambda$ & $\eta$ & $\tau_\eta$ & $L$ & $\tau_{L}$ & $Re_\lambda$ \\[2pt]
				\hline \\[-6pt]
				
				256 & 0.001 & 0.101 & 0.568 & 0.219 & 0.010 & 0.099 & 0.615 & 1.052 & 124 \\
			\end{tabular}
		\end{ruledtabular}
	\end{table}
	
	\subsection{Dynamics of Particles}
	\label{Par_des}
	
	
	Consistent with the assumptions described above, the dispersed phase consists of monodisperse particles (monomers) with a diameter $d = 1.1\times10^{-3}$, corresponding to $d/\eta = 0.11$. The carrier-fluid density is fixed at $\rho = 1.29$, while the particle density $\rho_p$ is varied to obtain different Stokes numbers at constant diameter. Across all cases, the density ratio ranges from $\rho_p/\rho \approx 150$ to $800$, which is consistent with the heavy-particle regime ($\rho_p/\rho \gg 1$).
	
	As discussed above, gravitational settling and inter-particle hydrodynamic interactions are neglected in the present study in order to isolate the effects of particle inertia and collision history. Under these conditions, the Maxey--Riley equation \cite{Maxey1983} governing particle motion may be reduced, to leading order, to the Stokes drag model:
	\begin{equation}
		\frac{d\mathbf{v}}{dt} = \frac{\mathbf{u} - \mathbf{v}}{\tau_p},
		\label{eq:particle_motion}
	\end{equation}
	where $\mathbf{v}$ is the particle velocity, $\mathbf{u}$ is the fluid velocity at the particle position. The fluid velocity at the particle position is obtained using linear interpolation from the surrounding grid nodes, and particle trajectories are advanced using a second-order Runge--Kutta scheme with exponential integrators, providing accurate simulation of particle trajectories even when $\tau_{p}$ is small\cite{Ireland_collins}. The parameter $\tau_p$ is the particle inertial response time, defined as:
	\begin{equation}
		\tau_p = \frac{1}{18} \left(\frac{\rho_p}{\rho} - 1\right) \frac{d^2}{\nu},
		\label{eq:response_time}
	\end{equation}
	where $\rho_p$ and $\rho$ are the density of particles and fluid respectively, $d$ is the particle diameter, and $\nu$ is the fluid kinematic viscosity.
	The Stokes number, which characterizes the importance of particle inertia in controlling the particle's motion in turbulence, is defined as the ratio of the particle response time to the Kolmogorov time scale: $St = \tau_p/\tau_\eta$, where $\tau_\eta$ is the Kolmogorov time scale given in Table~\ref{tab:fluid_params}.
	We keep the particle number density low enough such that collisions involving more than two particles simultaneously within the same simulation time step are rare, with the fraction of three-or-more-particle collisions among all collisions remaining in the range of $10^{-4}$ to $10^{-3}$.
	
	
	We systematically investigate the influence of Stokes number on collision and clustering statistics. To facilitate a meaningful comparison across different $St$, the number of particles ($N$) in each $St$-case is adjusted 
	 \fs{so that the average number of collisions per simulation time step satisfies $\langle \Gamma_{\mathrm G}\rangle\Delta t = 5 \text{--} 6$}, where this average is obtained from the ghost-particle simulations.
	This choice ensures comparable per-particle rare-collision condition across the series of simulations with varying Stokes numbers, with a normalized collision rate, $\Gamma_{\mathrm G}\tau_\eta/N$, of order $10^{-4}$--$10^{-5}$.
	Table~\ref{tab:particle_runs} lists the corresponding simulation parameters, grouped into two categories: simulations with varying Stokes numbers at an approximately constant collision rate, and simulations at fixed $St = 0.5$ with varying particle numbers.
	The $St=0.5$ case is adopted as the reference case throughout this work, as it represents the intermediate-inertia regime and serves as a reference linking the Stokes-number series and the particle-number series.
	
	
	\begin{table}[htbp]
		\caption{Parameters of the simulation runs: Stokes number ($St$), number of particles ($N$), and the mean collision count \fs{$\langle \Gamma_{\mathrm{G}} \rangle \Delta t$}  measured per simulation time step. The runs are grouped into two series, including simulations with varying Stokes numbers and others with varying particle numbers at a fixed Stokes number of 0.5.}
		\label{tab:particle_runs}
		\renewcommand{\arraystretch}{1.2}
		\centering
		\begin{ruledtabular}
			\small
			(a) Stokes number series (\fs{$\langle\Gamma_{\mathrm{G}}\rangle\Delta t = 5 \text{--} 6$})\\[3pt]
			\begin{tabular}{l *{10}{c}}
				$St$ & 0.01 & 0.05 & 0.10 & 0.30 & 0.50 & 0.70 & 0.90 & 1.00 & 2.00 & 3.00 \\
				$N$ ($10^6$) & 13.5 & 12.0 & 10.0 & 4.50 & 2.50 & 1.85 & 1.60 & 1.50 & 1.30 & 1.25 \\
				\fs{$\langle\Gamma_{\mathrm{G}}\rangle\Delta t$} & 5.72 & 5.63 & 5.61 & 5.84 & 5.77 & 5.61 & 5.59 & 5.43 & 5.63 & 5.70 \\
			\end{tabular}
			\\[6pt]
			(b) Particle-number series ($St = 0.5$)\\[3pt]
			\begin{tabular}{l *{5}{c}}
				$N$ ($10^6$) & 2.50 & 3.00 & 5.00 & 7.50 & 10.0 \\
				\fs{$\langle\Gamma_{\mathrm{G}}\rangle\Delta t$} & 5.77 & 7.68 & 21.38 & 48.25 & 86.95 \\
			\end{tabular}
		\end{ruledtabular}
	\end{table}
%
	
	\section{SIMULATION SETUP AND STATISTICAL ANALYSIS}\label{sec:postprocessing} 
	\subsection{Initialization}
	To ensure a fair and reproducible comparison, all cases were initialized from an identical particle configuration.
	A statistically stationary state of the carrier flow was attained after approximately $10\tau_L$, as indicated by the convergence of key turbulence parameters (e.g., dissipation rate and root-mean-square velocity) into a stable oscillatory regime. 
	Subsequently, non-colliding ghost particles were randomly seeded into the flow and allowed to disperse for approximately {$5\tau_L$ ($\sim 50\tau_\eta$)}, to achieve a well-mixed particle distribution. 
	However, when the CR and CN simulations transition from ghost to coalescing particles phase of the simulation, the first time step after the collision--coalescence subroutine is activated always contains numerous overlapping particle pairs.
	This would produce artificial burst of collisions, with instantaneous collision rates exceeding the long-time stationary value by orders of magnitude. 
	We thus exclude this transitional time step from the statistical analysis. Because these events lead to substantial particle (monomer) removal in the CN and CR simulations, we also remove the same set of particles from the ghost-particle simulations. 
	The resulting particle positions and velocities were then used as the common starting configuration for all simulations (GP, CN and CR systems). This removes any potential bias due to different initial particle configuration.In addition, all simulations were performed using the identical turbulent flow realization, with the carrier-phase velocity field evolving in exactly the same manner across different particle systems.
	
	Following this initialization, three different types of systems were simulated: the ghost-particle system; collision--coalescence system with continuous particle replenishment (CR) to maintain $N$ approximately constant; and collision--coalescence system without replenishment (CN). This allows us to investigate the influence of coalescence as well as particle replenishment on collision statistics compared to the idealized ghost-particle scenario. Each simulation was run for a duration of $20\tau_{L}$ ($\sim 200\tau_\eta$). In the GP cases, information of particle pairs with spatial separation $r \leq d$ (i.e., colliding) were recorded at every time step. This include the particle identifiers, positions, and velocities.

	In the CN and CR cases, each collision leads to coalescence, i.e., formation of larger particle, conserving mass and momentum. However, for the sake of simplicity and analytical clarity, the current work focuses on the dynamical behavior monodisperse systems only, as a result, larger particles are removed from the simulations once they are formed. A full study of the effects of polydispersity is beyond the scope of this work (limited evaluation of its effect will be presented where necessary).

	In the CR case, particles removed due to coalescence were replenished via regular injection of new particles at a constant rate, allowing $N$ to remain close to that in the ghost particle case. Throughout the simulations, maximum deviation in $N$ for low-inertia particles ($St \le 1$) remains below 1\%; while for larger particles this reach a maximum of 2\% (for details, see Appendix~\ref{app:injfreq}). 
		
	\subsection{Computation of RDF}
	\label{sec:rdf}
	 
	To quantify the small-scale spatial distribution of particle pairs, the RDF is computed following:
		\begin{equation}
			\label{eq:rdf}
			g(r) = \frac{\psi(r)/N}{(N-1)\,\delta V_r / V},
		\end{equation}
	where $\psi(r)$ is the total count of particle pairs whose separation lies within $[r, r+\delta r]$ (double counting is practiced), $V$ is the domain volume, and $\delta V_r = 4\pi r^2 \delta r$ is the spherical-shell volume used for normalization. The RDF was computed using particle data sampled every $2.5\tau_\eta$.
		
	Using the above formula, we evaluate the RDF for three cases: $g_{\mathrm{G}}(r)$, $g_{\mathrm{CR}}(r)$, and $g_{\mathrm{CN}}(r)$, corresponding to the conventional ghost-particle simulation and the collision--coalescence simulations with and without particle replenishment. 
	For the ghost-particle case, we further separate the RDF conditioned on the sign of the radial relative velocity $w_r$, defining the approaching-pair RDF $g_{\mathrm{G}}^{\scriptscriptstyle (-)}(r) \equiv g_{\mathrm{G}}(r\,|\, w_r<0)$ for particle pairs moving toward each other. Conditioning on $w_r<0$ isolates the subset of pairs that are kinematically in collision courses. 
	Conversely, $g_{\mathrm{G}}^{(+)}(r)$ could be defined for separating pairs in a complementary manner.
	
	The RDF at particle contact is represented using the shorthand $g(d)$, which formally corresponds to the limiting value of the RDF as the interparticle separation approaches the particle diameter from above:
		\begin{equation}
			g(d) \equiv \lim_{r \to d^+} g(r).
		\end{equation}
	This is because the RDF is strictly zero for $r \le d$ as the simulations assume that coalescence occurs instantaneously\footnote{More realistic treatment that introduces an extra coalescence time scale is possible but beyond our scope.}.
	In practice, $g(d)$ is estimated by averaging $g(r)$ over a thin spherical shell immediately above particle contact ($r = d$).	
	
	\subsection{Collision Analysis}
	\label{sec:collision}
	In the ghost-particle simulation, the analysis focuses on correlated successive collisions, defined as successive collision events experienced by a particle that occur within a characteristic correlation timescale. It is worth noting that successive collision does not necessarily imply repeated collision between the exact same particle pair, i.e., a particle may collide with different partners in the subsequent collision. 
		
	To this end, collision statistics are extracted through a post-processing procedure that
	reconstructs discrete collision events from the raw contact records.
	For each particle pair, contact times are first sorted chronologically. 
	Consecutive contact records in adjacent simulation time steps are regarded as belonging to the same collision event, representing continuous physical contact between the two particles. A new collision event is identified only after the contact has been interrupted for at least one simulation time step. 
	Based on the reconstructed collision-event sequences
	, collision rates and related statistics are computed for subsequent statistical analysis.
		
	The collision rate $\Gamma$ is defined as the number of collision events occurring per unit time. 
	For DNS with periodic boundaries such as ours, the corresponding collision kernel is canonically defined as:
	\begin{equation}
			\label{eq:K_def}
			K = \frac{2 \Gamma V}{N^2},
	\end{equation}
	where $V$ is the simulation domain volume and the factor "$2$" arises from (unordered) pair-combinatorics. In the context of Smoluchowski population balance equations, for monodisperse particles, $K/2$ multiplied by particle density squared gives the per unit volume collision rate; while for collision rate between particle of different size/type, $K$ instead of $K/2$ is used.
	\section{Heuristic Discussion of an Earlier Model}\label{sec:prelude}
	
	In this section, we discuss a previously presented simple model that is of significant relevance to the current work in order to provide context for the analysis and findings to be presented  in the sequel.
	
	In an earlier work\cite{Saw2022}, a phenomenological model was presented to predict the mean radial velocity (MRV) of particle pairs in a system of coalescing particles in turbulence (equivalent to CR in this work). The core idea of the model rests on applying the central limit theorem on the approach (exit) angles of relative particle trajectories followed by angle-filtering consistent with the geometry of the coalescence physics. 
	The same model (and its principles) 
	is also applicable for the prediction of the monodisperse RDF. Following the same segregation based on the sign of the radial velocity and filtering approach, the RDF may be decomposed into contributions from positive and negative radial velocity events:   
	\begin{equation}
		g_{\text{model}}(r) = p_{\scriptscriptstyle -} \,g_{\text{G}}(r) \,+\, p_{\scriptscriptstyle +} \,g_{\text{G}}(r) \frac{1}{\int^{\frac{\pi}{2}}_{0} P(\theta)d\theta} \int^{\frac{\pi}{2}}_{\theta_m} P(\theta) d\theta, \ \  \  \  \	 \theta_m = \sin^{-1}(d/r),
		\label{Eqn_rdf_model}
	\end{equation}
	where the first (last) term in the sum corresponds to the contribution of particle pairs with negative (positive) radial velocities; $p_{\scriptscriptstyle \pm}$ are constant decomposition factors; $g_{\text{G}}(r)$ is the RDF of the equivalent ghost particle system; $P(\theta)$ is the probability distribution function of the angle between the relative velocity and displacement of a particle pair (exit angle); $d$ is the particle diameter. Essentially, the definite integral over $P(\theta)$ in the second term corresponds to the realizable exit angles under the geometric constraints of the coalescence physics\cite{Saw2022, Meng2026}. For the first term (negative velocities) the equivalent integral results in a value of unity since all angles are admissible. In the most basic form, the decomposition factors $(p_{\scriptscriptstyle -},\, p_{\scriptscriptstyle +})$ are set to $(1/2, \,1/2)$ implying that one assumes equal probability for positive and negative radial velocity events (which is equivalent to assuming that the distribution of radial velocity is not skewed). 
	However, it is well established that turbulent small-scale velocity difference statistics are negatively skewed (cf. Fig.~\ref{fig:n2} and Fig.~\ref{fig:n3}). 
	
	\begin{figure}[htbp]
		\centering
		\includegraphics[width=0.65\linewidth]{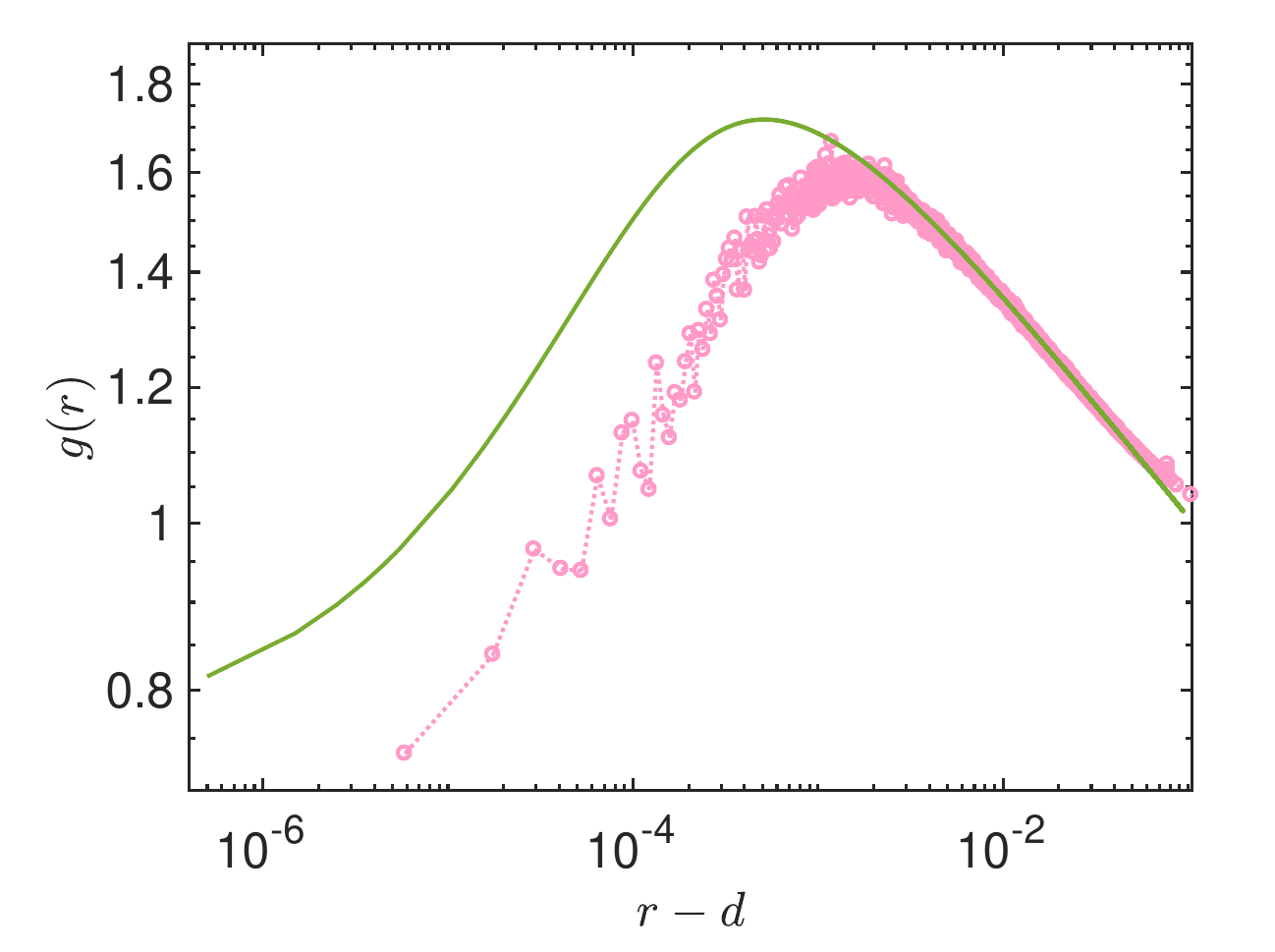}
		\caption{
			{Comparison of the radial distribution function predicted by the theoretical model with direct numerical simulation results for collision--coalescence (CR) simulations at $St=0.1$. The solid green line is the model curve and the circles represent the RDF from the collision--coalescence simulation with particle replenishment $g_{\mathrm{CR}}(r)$. The model captures the overall shape of $g_{\mathrm{CR}}(r)$, reflecting its ability to reproduce the dominant spatial features of collision--coalescence statistics. The current work aims to understand the main remaining discrepancy (details in text).}
		}
		\label{fig_teaser}
	\end{figure}
	
	In Figure~\ref{fig_teaser}, the solid green line represents the prediction of Eq.~(\ref{Eqn_rdf_model}) for the case of $St=0.1$, using the more realistic decomposition factors $(p_-, ~p_+) = (0.4, 0.6)$, which are consistent with our DNS findings for the ghost-particle system (note also that $p_-$ ($p_+$) is essentially the probability of sampling negative (positive) radial velocities). The   {$g_{\text{G}}(r)$} used here is a power law $c_0 \, r ^{-c_1}$ with constants fitted to the corresponding ghost-particle DNS outcome; alternatively, the constants may also be obtained theoretically from e.g., Chun \textit{et al.}\cite{Chun2005} for the Stokes number in question. A central-limit model of $P(\theta)$ is given by Saw and Meng\cite{Saw2022}; however, to facilitate the analytical evaluation of the integrals in the equation, here we approximate $P(\theta)$ using a Gaussian distribution of $(\mu,~\sigma) = (\frac{\pi}{2},~ 0.433)$, which we observed to closely reproduce the original model in spite of minor deviations at the tails. We compare the model prediction with the RDF resulting from the DNS of the coalescing system (CR) with the same $St$ value. Fig.~\ref{fig_teaser} shows that the model closely reproduces the qualitative trend of the RDF of the coalescing system.  However, the model curve exhibits a vertical shift relative to the DNS result, differing from the latter by a multiplicative factor of order unity. This discrepancy persists alongside an inaccurate lateral peak position. We are primarily concerned with the former shortcoming, as it would lead to an overestimation of particle collision rate---which is generally assumed to be proportional to the value of RDF at contact\cite{Sundaram1997} $g(d)$. 
	
	The interest to understand and fill the theoretical gap represented by the discrepancy described above is the main motivation for the core analysis in the current work. The findings presented here shall be used to inform subsequent theoretical work to fill this gap.
	
	\section{RESULTS AND DISCUSSION}\label{sec:res}	
	\subsection{Results on Radial Distribution Function and Collision Kernel}
	\label{sec:discrepancy}
	
\begin{figure}[htbp]
			\centering
			\includegraphics[width=1\linewidth]{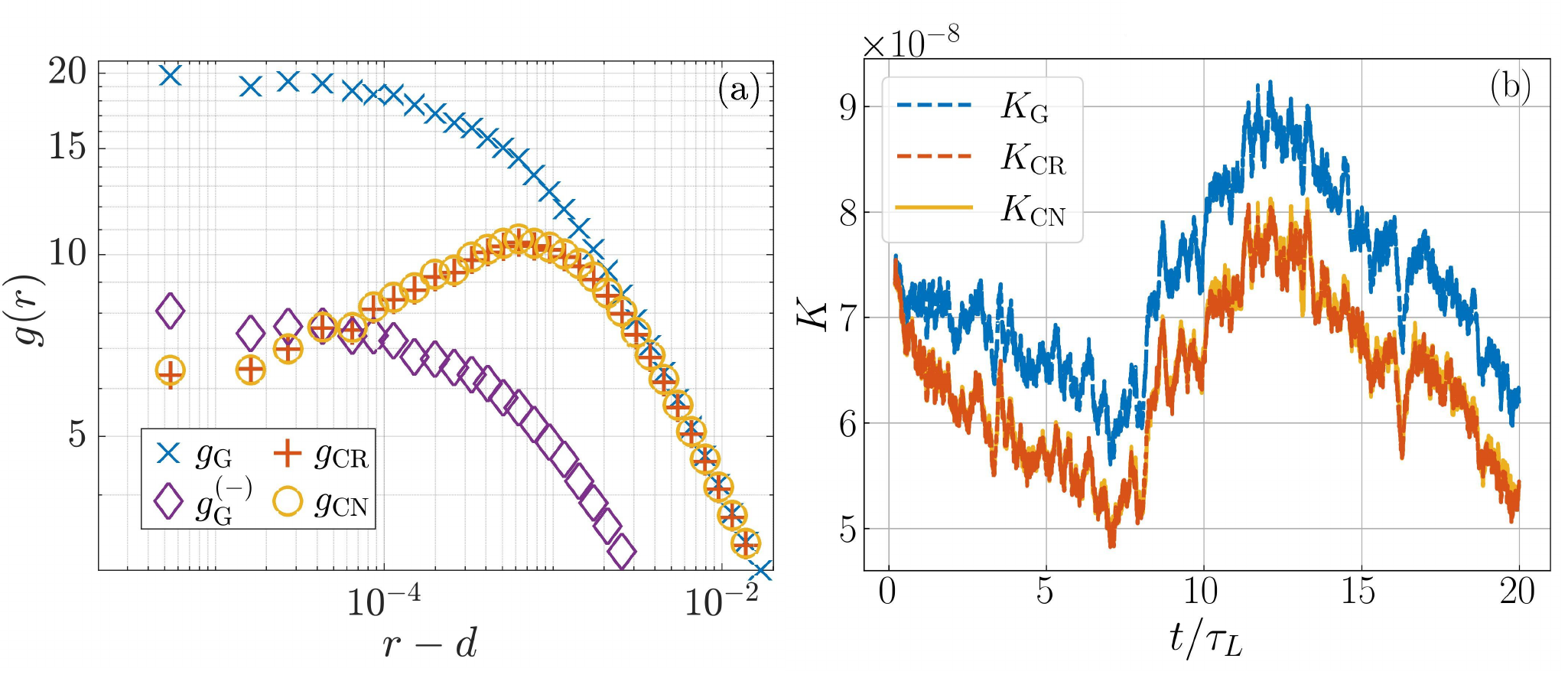}
			\caption{
				(color online) Comparison between the ghost-particle and collision--coalescence simulations for $St = 0.5$ and $N = 2.5\times10^{6}$ (for CN, $N$ starts at $2.5\times10^{6}$ and decreases with time).
				(a) Radial distribution functions, which include the ghost-particle result $g_{\mathrm{G}}(r)$ (blue crosses), the ghost-particle RDF conditioned on negative radial relative velocity $g_{\mathrm{G}}^{\scriptscriptstyle (-)}(r)$ (purple diamonds), the case of collision--coalescence with particle replenishment $g_{\mathrm{CR}}(r)$ (red + symbol), and without particle replenishment $g_{\mathrm{CN}}(r)$ (yellow circles).
				(b)\fs{*Have modified y-axis value*} Temporal evolution of the collision kernel ($K$). From the top, the ghost-particle $K_{\mathrm{G}}$, the collision--coalescence cases $K_{\mathrm{CR}}$ and $K_{\mathrm{CN}}$ corresponding to systems with and without particle replenishment respectively.  
			}
			\label{fig:n2}
		\end{figure}

	We begin with two statistics---the radial distribution function, $g(r)$, which quantifies the spatial clustering of particles, and the collision kernel, $K$, an intrinsic metric of the collision process. 
	As discussed in Section~\ref{Par_des}, we use the reference case $St = 0.5$, $N=2.5\times10^6$ (Fig.~\ref{fig:n2}) to illustrate the main findings. All simulations are performed under rare-collision conditions, as detailed therein. This intermediate-inertia case 
	exhibits relatively large discrepancies between the ghost-particle and collision--coalescence systems, making it a representative example for examining the differences between the two systems.
	
	Figure~\ref{fig:n2}(a) shows the RDF plotted as a function of $r-d$ on a log-log scale to highlight the small scale behavior. In addition to the conventional ghost-particle RDF, $g_{\mathrm{G}}(r)$, we also show the ghost-particle RDF conditioned on approaching particle pairs, $g_{\mathrm{G}}^{\scriptscriptstyle (-)}(r) \equiv g_{\mathrm{G}}(r\,|\, w_r<0)$, which only counts the particle pairs that have negative radial relative velocity $(w_r<0)$ when evaluating the numerator of the RDF formula (c.f. Eq.~(\ref{eq:rdf})). The RDFs from collision--coalescence simulations with and without particle replenishment are included for comparison ($g_{\mathrm{CR}}(r)$ and $g_{\mathrm{CN}}(r)$ respectively).  
	Fig.~\ref{fig:n2}(a) shows that the RDFs of the three systems are nearly identical at larger separations and begin to diverge only in the near-contact region.
	As the interparticle separation approaches the contact distance ($r \to d$), the ghost-particle RDF $g_{\mathrm{G}}(r)$ increases monotonically toward a finite contact value. 
	This near-contact deviation is found to occur at a similar separation scale in the other Stokes numbers considered, at approximately $r-d \sim 2d$. 
	A local maximum is also visible in $g_{\mathrm{CR}}$ and $g_{\mathrm{CN}}$ just outside the near-contact depletion region consistent with the sharp depletion previously reported by Meng \&	Saw~\cite{Meng2023}. 
	The conditional RDF $g_{\mathrm{G}}^{\scriptscriptstyle (-)}(r)$, however, is uniformly downward shifted, such that near $r=d$, $g_{\mathrm{G}}^{\scriptscriptstyle (-)}(r)$ becomes close to $g_{\mathrm{CR}}(r)$ and $g_{\mathrm{CN}}(r)$. This suggests that, to the lowest order, $g_{\mathrm{G}}^{\scriptscriptstyle (-)}(d)$ is a good estimate of $g_{\mathrm{CR}}(d)$, opening the possibility of predicting the RDF and collision rate of coalescing systems via the well studied ghost-particle statistics. 
	
	We also see that $g_{\mathrm{CR}}(r)$ and $g_{\mathrm{CN}}(r)$ remain nearly indistinguishable throughout the whole range of $r$, consistent with the fact that the RDF, by definition, is number density (squared) normalized and thus insensitive to variations in particle number density.
	The temporal evolution of the corresponding collision kernels is shown in Fig.~\ref{fig:n2}(b). 
	The kernel is computed from collision counts following the procedure described in Section~\ref{sec:collision}, and is related to the collision rate $\Gamma$ through Eq.~(\ref{eq:K_def}). To suppress short-time fluctuations, a moving average with a window of 200 time-steps was applied.  Consistent with the RDF behavior, the ghost-particle kernel $K_{\mathrm{G}}$ remains systematically larger than those from the collision--coalescence simulations. In contrast, the latter kernels, $K_{\mathrm{CR}}$ and $K_{\mathrm{CN}}$, are nearly indistinguishable, indicating that the collision kernel statistics are insensitive to the variation in particle number density under the present conditions.
		
\begin{figure}[htbp]
			\centering
			\includegraphics[width=0.55\textwidth]{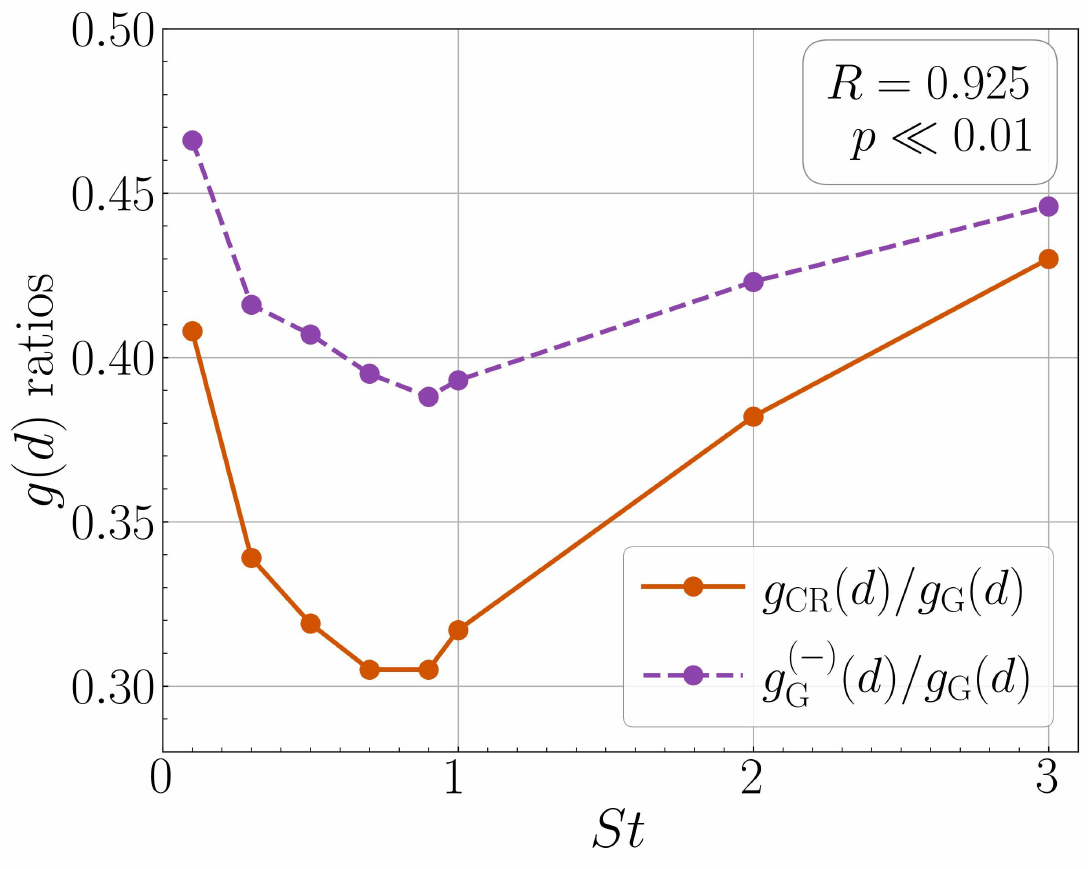}
			\caption{Comparison at particle contact ($r=d$) between the approaching-pair fraction in the ghost-particle simulations,
			$g_{\mathrm{G}}^{\scriptscriptstyle (-)}(d)/g_{\mathrm{G}}(d)$ (top plot), and the at-contact RDF value from the collision--coalescence simulations divided by the corresponding ghost-particle value, $g_{\mathrm{CR}}(d)/g_{\mathrm{G}}(d)$ (bottom plot), across different Stokes numbers. 
			The Pearson correlation coefficient between the two plots and the associated p-value are $R = 0.925$ and $p \ll 0.01$, respectively.}
			\label{fig:n3}
	\end{figure}

	Since collisions are governed by particle-pair statistics at contact, we are especially interested in the value of $g(d)$. Figure~\ref{fig:n3} compares the fraction of approaching particle pairs in the ghost-particle simulations, $g_{\mathrm{G}}^{\scriptscriptstyle (-)}(d)/g_{\mathrm{G}}(d)$, with the ratios between 
	$g(d)$ from the collision--coalescence simulation and that from the corresponding ghost-particle simulation, $g_{\mathrm{CR}}(d)/g_{\mathrm{G}}(d)$. Across the range of Stokes numbers considered, $g_{\mathrm{G}}^{\scriptscriptstyle (-)}(d)/g_{\mathrm{G}}(d)$ lies between approximately $0.38$ and $0.47$. This indicates that there are always more separating particle pairs than approaching ones, consistent with the experimentally observed negative skewness of the probability distribution of particle relative velocity\cite{Saw2014}. For a system of particles in (statistical) steady state, negative skewness, i.e., the tail of the distribution for negative values is more stretched, implies that the negative velocity events (approaching pairs) occur with nominally larger speeds than the positive (separating) events. At steady state, however, the positive events should occur more frequently so that the combined average relative velocity is zero. Such negative skewness is a direct consequence of particle inertia and was found to increase with $St$ for $St \lesssim 0.5$ in Saw \textit{et al.} (2014)\cite{Saw2014}. The current results, however, suggests that this skewness becomes gradually weaker for heavier particles of $St \gtrsim 1$.  
	
	We also see that the ratio $g_{\mathrm{CR}}(d)/g_{\mathrm{G}}(d)$ exhibits a strong linear correlation with the approaching-pair fraction $g_{\mathrm{G}}^{\scriptscriptstyle (-)}(d)/g_{\mathrm{G}}(d)$, with a Pearson correlation coefficient of  $R = 0.925$. This close correlation suggests that $g_{\mathrm{G}}^{\scriptscriptstyle (-)}(d)$ is a decent proxy for estimating corresponding collision statistics in coalescing particle systems.
	
	\subsection{Discrepancy Analysis: Velocity-Filtered Ghost-Particle Statistics vs. Coalescing Systems.}
	
	Building on the observation in the preceding section that $g_{\mathrm{G}}^{\scriptscriptstyle (-)}(d)$ systematically overestimates $g_{\mathrm{CR}}(d)$, we now examine the dependence of this discrepancy on Stokes number and particle density. 
As shown earlier, $g_{\mathrm{G}}^{\scriptscriptstyle (-)}(d)$ captures the subset of particle pairs on a collision course and correlates closely with $g_{\mathrm{CR}}(d)$. This conditional quantity provides a link between ghost-particle and coalescing-system statistics: it is readily obtained from standard ghost-particle simulations and is directly related to the heuristics theoretical descriptions of approaching particle pairs in Section \ref{sec:prelude}.
To examine the remaining discrepancy between this quantity and coalescing-system statistics in a systematic manner, we compare the normalized RDF and collision-kernel ratios shown in Fig.~4, using the ghost-particle results as a common reference.
	The left panel shows their dependence on the Stokes number ($St$) under comparable collision  \fs{frequency, with $\langle\Gamma_{\mathrm{G}}\rangle\Delta t = 5\text{--}6$}. We first consider the ratio of the RDFs at contact, $\langle g_{\mathrm{CR}}(d) \rangle / \langle g_{\mathrm{G}}^{\scriptscriptstyle (-)}(d) \rangle$. This ratio remains below unity across the entire range of $St = 0.01\text{--}3.0$, showing again that the ghost-particle simulation, even when only accounting for approaching pairs, overestimates the at-contact particle concentration of a coalescing system. 
	The ratio exhibits a non-monotonic dependence on $St$, reaching a minimum around $St \approx 0.7$. This indicates that the deviation between ghost-particle and coalescing-system RDFs is most pronounced in the intermediate-inertia regime ($St \sim O(1)$), where both preferential concentration and inertial decorrelation influence the formation of near-contact particle pairs.
	\begin{figure}[htbp]
			\centering
			\includegraphics[width=0.98\linewidth]{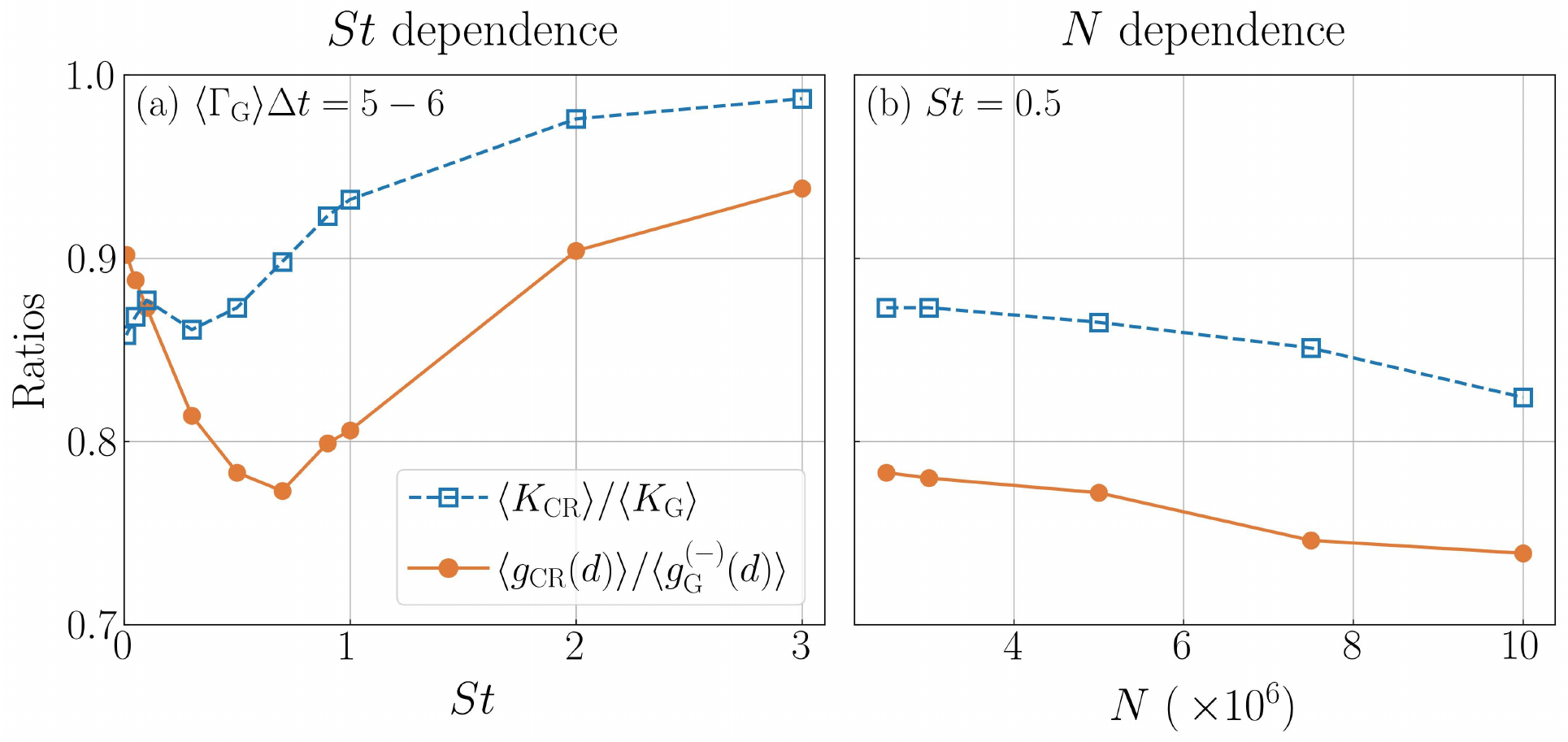}
			\caption{
				(color online) The systematic overestimations by the ghost-particle simulation relative to the coalescing particles system, and their dependence on Stokes number ($St$) and number of particles ($N$). 
				The two ratios plotted are: the collision kernel ratio 
				$\langle K_{\mathrm{CR}} \rangle / \langle K_{\mathrm{G}} \rangle$ (blue squres) and the RDF ratio at contact
				$\langle g_{\mathrm{CR}}(d) \rangle / \langle g_{\mathrm{G}}^{\scriptscriptstyle (-)}(d) \rangle$ (orange circles), where $g_{\mathrm{G}}^{\scriptscriptstyle (-)}(d)$ denotes the ghost-particle RDF value at contact conditioned on approaching particle pairs. 
				(a) Dependence on Stokes number under comparable collision \fs{frequency,
				with$\langle\Gamma_{\mathrm{G}}\rangle\Delta t = 5\text{--}6$}. 
				(b) Dependence on the $N$ at $St = 0.5$. }
			\label{fig:n4}
	\end{figure}
		
	In the limit of small Stokes number (the tracer-like regime), the ratio increases to around $0.9$. 
	At larger $St$, the ratio exhibits a non-monotonic dependence on $St$, reaching a minimum of approximately 0.77 at $St \approx 0.7$, suggesting that the influence of correlated successive collisions, if it is indeed the dominant cause of this discrepancy, is strongest in this Stokes-number range. 
	Beyond that, the ratio gradually recovers towards unity as particle motion becomes increasingly decorrelated from small-scale turbulent structures, suggesting that correlation with turbulent eddies is a key source of the phenomenon.
		
	Turning to the collision kernel, we examine the ratio $\langle K_{\mathrm{CR}} \rangle / \langle K_{\mathrm{G}} \rangle$, which likewise remains below unity over the entire $St$ range, indicating a systematic overestimation of collision rates by the ghost-particle simulations. Its dependence on $St$ is also non-monotonic, reaching a minimum of approximately $0.86$ at $St \approx 0.3$. 
	At intermediate inertia ($St \gtrsim 0.1$), the trends of the kernel ratio and the RDF ratio remain closely coupled, indicating that discrepancies in the collision kernel in this regime are primarily driven by differences in the RDFs. In contrast, at the lower-inertial regime ($St \lesssim 0.1$), a clear decoupling emerges. Although the at-contact RDF ratio increases as $St \to 0$, the kernel ratio does not follow. 
	Since the leading-order collision kernel scales as $K \sim g(d)\langle w_r^- \rangle$ (see e.g., Sundaram \& Collins\cite{Sundaram1997}), the kernel discrepancy at very small $St$ is likely dominated by the relatively stronger variation in the radial velocity statistics. 
		
	Beyond $St \approx 0.3$, the kernel ratio increases and asymptotically approaches unity for $St \gtrsim 2$. This trend reflects an increasing resemblance to a kinetic system of gas molecules. In this regime, successive collision events are free from any carrier flow induced correlation.
	The kernel-ratio trends seen here are consistent with earlier studies which showed that ghost-particle simulations systematically overestimate collision rates due to the neglect of correlated particle dynamics and multiple collisions (Vo{\ss}kuhle \textit{et al.} \cite{Vokuhle2011, Vokuhle2013}).
		
	The relationship between these two ratios can be understood through the kinematic framework of Sundaram and Collins \cite{Sundaram1997},  which claims that the collision kernel is proportional to the product of the RDF and the mean approaching radial relative velocity at contact i.e., $K \propto g(d)\langle w_r(d) ~ | w_r<0 \rangle$. In this view, the RDF ratio being systematically below unity implies the same for the kernel, as we observe here. The offset between the locations of the minima in the two ratios indicates that variations in the mean radial relative velocity partially also play a significant role. 
		
	In contrast to the strong systematic dependence on $St$, the overestimation shows only a weak, secondary dependence on the particle number $N$. As shown in Fig.~\ref{fig:n4}(b) for $St=0.5$, both ratios exhibit a very slight decreasing trend with increasing $N$.
	Since the ratios are normalized by ghost-particle statistics, the leading-order scaling with $N$ is removed by construction. Although a weak residual dependence on $N$ is observed, its physical origin is not yet understood. This remains an open question for future work.
	The magnitude of this variation remains small compared with the changes observed across the Stokes-number range in Fig.~\ref{fig:n4}(a).
	Similar weak $N$-dependence is also observed for the cases of $St=0.1$ and $St=1.0$. 
		
	\subsection{Impact of Correlated Successive Collisions on Particle Collision Statistics}
	The systematic discrepancy identified above, whereby $g_{\mathrm{CR}}(d)$ is consistently smaller than $g_{\mathrm{G}}^{\scriptscriptstyle (-)}(d)$ and similarly for the kernels, requires a physical explanation in terms of particle collision dynamics. We consider the possibility that this discrepancy can be attributed to the role of collision memory or history. We define the concept of correlated collisions as collision events for which at least one of the participating particles has experienced a prior collision. Such correlations naturally arise in the ghost-particle approach but are inherently absent in collision--coalescence simulations, where particles merge after colliding to form a new larger particle (i.e., the original particles cease to exist).
		
	\begin{figure}[htbp]
		\centering
		\includegraphics[width=0.55\linewidth]{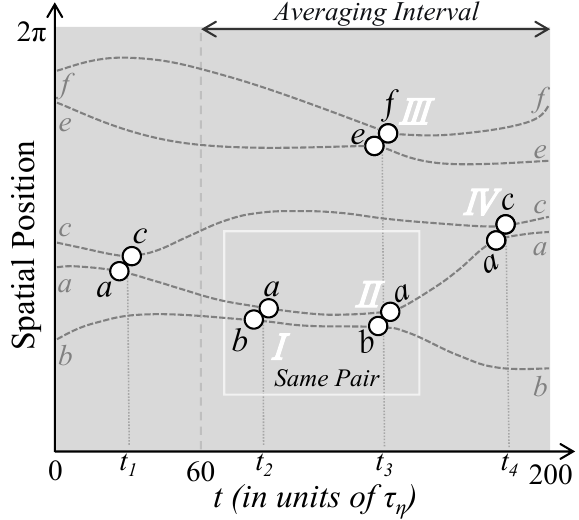}
		\caption{
		Schematic illustrating the different cases associated with collision history. The horizontal axis represents time, and the vertical axis represents a generalized spatial coordinate. Dashed lines indicate particle trajectories.
		Collision events are marked by Roman numerals (I$\text{--}$IV). For a given collision event, 
		a backward trace is performed through the histories of both particles to identify the last collision involving either particle. 
	     Four distinct cases are illustrated. 
		 Case I: Detected collision $(a,b)$ at $t_2$; particle $a$ had prior collision $(a,c)$ at $t_1$. 
		 Case II: Same pair $(a,b)$ re-collides at $t_3$ after initial collision at $t_2$. 
		 Case III: Collision of $(e,f)$ at $t_3$, where both particles had no prior collision history. 
		 Case IV: Collision of $(a,c)$ at $t_4$, where both particles had prior collision histories ($a$ at $t_3$ and $c$ at $t_1$). 
		 For statistical analysis, only the final $140\tau_\eta$ of the record is used as the averaging interval, ensuring that the backward trace history for all classified collision events extends to at least $60\tau_\eta$.
		}
		\label{fig:n5}
		\end{figure}
		
		To ensure reliable statistics, we restrict our analysis to collisions occurring within the final $140\tau_\eta$ ($\approx14\tau_{L}$) of the simulation, as illustrated in Fig.~\ref{fig:n5}.  
		This interval is chosen to guarantee that the backward trace for each considered collision extends in the GP system at least $60\tau_\eta$, ensuring that sufficient trajectory history is available for identifying prior collision events. For consistency, the GP and CN statistics are evaluated over the same sampling interval.
		For each collision occurring at time $t$, the trajectories of both particles are traced backward in time over the interval $(0,\,t)$ to identify prior collision events.
		A colliding particle pair is classified as having a collision history if at least one of the two particles involved experienced an earlier collision within the history tracing window.
		This definition includes both past collisions of the same particle pair and collisions involving different partners,  thereby capturing the broader class of correlated successive collisions in the ghost-particle framework. 
		This requirement was verified to be sufficient for capturing the relevant collision history while maintaining adequate statistical sampling.
		Four illustrative cases in Fig.~\ref{fig:n5} demonstrate the different scenarios: Case I involves a colliding pair where one of the particles has a collision history; Case II shows the re-collision of the same pair; Case III is a collision where both particles have no collision history; and Case IV presents a collision where both particles have distinct collision histories.
		
		\begin{figure}[htbp]
			\centering
			\includegraphics[width=1\linewidth]{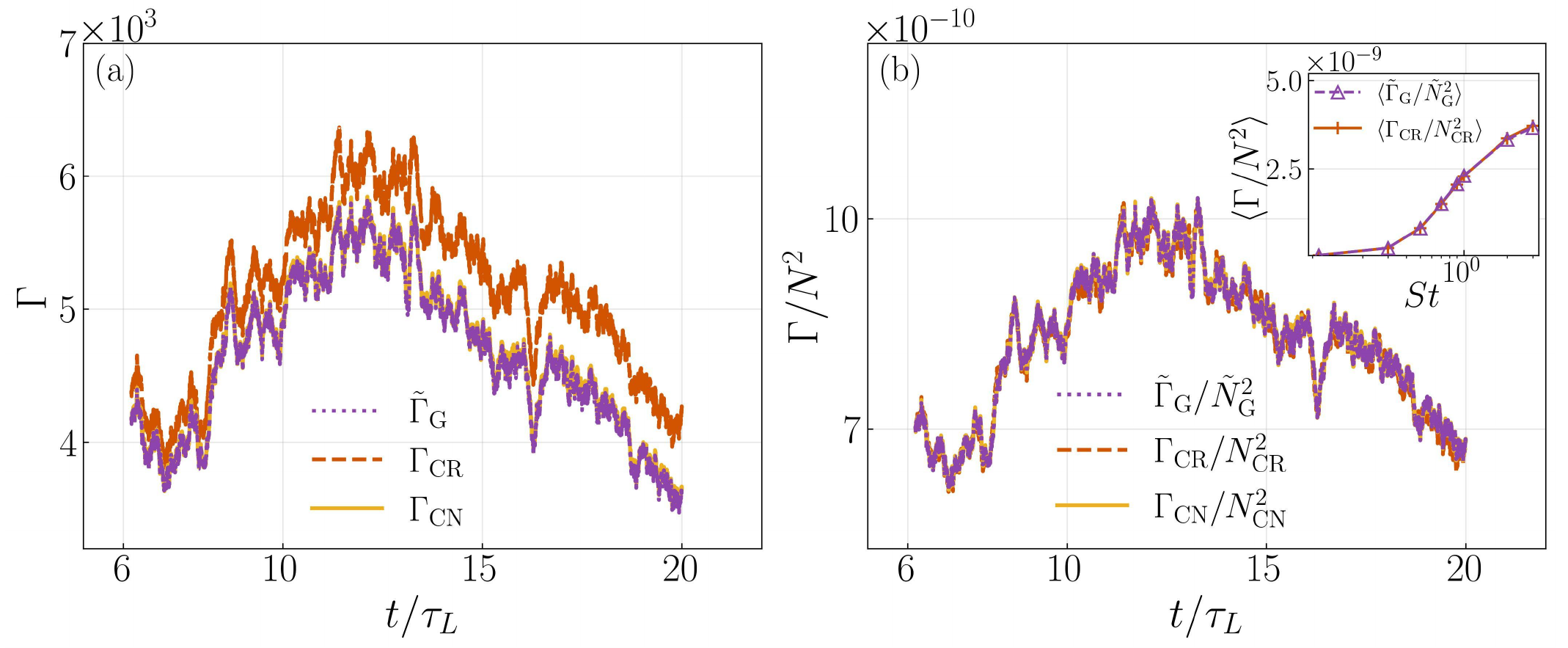}
			\caption{\fs{*Have modified y-axis value, and reduced the font size of inset*} 
			Time evolution and Stokes number dependence of collision statistics for the representative case of $St=0.5$ and initial number of particles $N(t=0)=2.5\times10^6$.
			\textbf{(a)} Collision rate $\Gamma$ as a function of $t/\tau_{L}$, comparing the 
			history-filtered ghost-particle rate $\tilde{\Gamma}_{\mathrm{G}}$, the collision--coalescence case with replenishment (${\Gamma}_{\mathrm{CR}}$), and without replenishment (${\Gamma}_{\mathrm{CN}}$). 
			\textbf{(b)} Time series of the reduced kernel $\Gamma/N^{2}$ for the same three cases. \textbf{Inset)} Time-averaged kernels $\langle \Gamma/N^{2} \rangle$ of the ghost-particle and CR systems, at various values of $St$. 
			In all systems, $\Gamma$ is normalized by $N^2$ at each time step before averaging, ensuring a consistent comparison. 
			}
			\label{fig:n6}
		\end{figure}
		
	Guided by Fig.~\ref{fig:n5}, we now demonstrate the role of correlation between successive collisions as a potential mechanism for the observed discrepancy between the ghost-particle and collision--coalescence simulations.
	We define the history-free (i.e., first-time) collision rate, $\tilde{\Gamma}_{\mathrm{G}}$,  as the ghost-particle collision rate conditioned on neither of the two participating particles having experienced a prior collision since the effective start of the simulation (excluding spin-up stages). 
	This quantity is therefore free of any effects of 
	correlated successive collisions and its magnitude typically decays with time as the number of eligible particles in the system progressively declines; this is in contrast to the conventional ghost-particle rate $\Gamma_{\mathrm{G}}$, which is statistically stationary.
	To assess how this history-conditioned or history-filtered collision rate compares with the rates in the collision--coalescence systems, we juxtapose their respective trends in Fig.~\ref{fig:n6}.
	
	In Fig.~\ref{fig:n6}, panel (a) shows the temporal evolution of the collision rates for the representative case ($St=0.5$ and $N(t=0) = 2.5\times10^{6}$), for all three systems. 
	We see that $\tilde{\Gamma}_{\mathrm{G}}$ closely follows $\Gamma_{\mathrm{CN}}$ at all times. 
	The relative difference $| \langle \tilde{\Gamma}_{\mathrm{G}}\rangle - \langle \Gamma_{\mathrm{CN}} \rangle |/\langle\Gamma_{\mathrm{CN}}\rangle$ 
	remains below approximately $1.5\%$ for all $St$, with $ \langle \tilde{\Gamma}_{\mathrm{G}} \rangle$ systematically but only marginally smaller than $ \langle \Gamma_{\mathrm{CN}} \rangle$. This corroborates the conjecture that possible correlations in successive collisions  could be the cause of the discrepancy between the collisional statistics of ghost and coalescence systems, in the sense that after experiencing a collision, a ghost particle will have an enhanced chance of colliding again, consistent with the findings of Vo{\ss}kuhle \textit{et al.}\cite{Vokuhle2013}.
	Another observation  is that both $\tilde{\Gamma}_{\mathrm{G}}$ and $\Gamma_{\mathrm{CN}}$ become consistently below $\Gamma_{\mathrm{CR}}$ over time. This is simply the consequence of the decreasing particle density (or $N$) over time in the CN and the history-filtered ghost system.
	
	A minor point worth noting is that, unlike in the CN simulations, in the ghost-particle simulation, particles that have previously collided are not removed from the system; they are simply not counted when computing $\tilde{\Gamma}_{\mathrm{G}}$. This has an undesired second order effect where these particles may collide with other history-free particles, thus further reducing the number of particles that are eligible to contribute to $\tilde{\Gamma}_{\mathrm{G}}$ over time relative to the CN system. As a consequence, $\tilde{\Gamma}_{\mathrm{G}}$ is slightly lower than its CN counterpart, despite the overall similarity between the two curves. This discrepancy will become more apparent if the particle density and likewise the collision rate is increased.
	
	To isolate the effect of particle density, Fig.~\ref{fig:n6}(b) shows the time evolution of the instantaneous reduced kernels $\Gamma/N^{2}$, which is equivalent to $K/(2V)$. 
	The inset presents the corresponding time-averaged reduced kernel. Here it is important to distinguish between systems with stationary and decaying $N$. For the ghost-particle and CR simulations, $N$ remains approximately constant in time, so that $\langle \Gamma \rangle / N^2$ and $\langle \Gamma / N^2 \rangle$ are very close in practice. In contrast, for the CN and history-filtered ghost systems, $N$ varies with time; therefore, the normalization is performed prior to averaging, i.e., the quantity shown is $\langle \Gamma / N^2 \rangle$. 
	This ensures a consistent comparison across the different systems. This result provides evidence that the discrepancy between ghost-particle and collision--coalescence statistics originates primarily from correlations between current collision probability and collision history, or in other words, a memory effect.
	
	\subsection{Generalized History-Filtering and Collision-Correlation Time}
	
	Motivated by the observations in the preceding section,  we now generalize the concept of history-filtered rate (as it relates to $\tilde{\Gamma}_{\text{G}}$).
	We introduce a memory time parameter $\tau$, such that the generalized history-filtered rate $\tilde{\Gamma}_{\text{G}}(t;\, \tau)$ is defined as the collision rate that counts only collisions where both participating particles have no collision history within the interval $(t-\tau,\,t)$. In this new picture, $\tilde{\Gamma}_{\text{G}}(t;\, \infty)$ is equivalent to the previously unparameterized $\tilde{\Gamma}_{\text{G}}$, which represents the strictly history-free rate. In the opposite limit, $\tilde{\Gamma}_{\text{G}}(t;\, 0)$ is simply ${\Gamma}_{\text{G}}$, the conventional ghost system collision rate. 
	
	Given this context, we can now postulate that there exists a timescale $\tau_{\mathrm{cc}}$, between the two limits ($ 0,\, \infty$), such that when $\tau \ge \tau_{\text{cc}}$, the history-filtered ghost-particle kernel reproduces,  to leading order, the kernel of the coalescing systems, namely,  
	\begin{equation}
		\frac{ \left\langle\, 
			\tilde{\Gamma}_{\mathrm{G}}^{}(t;\, \tau) 
			\,\right\rangle_{} } { {\left\langle\, \tilde{N}_{\text{G}}(t;\,\tau) \,\right\rangle}^2 }
			=
			\left\langle\frac{ \Gamma_{\mathrm{CN}}  } { { N_{\mathrm{CN}} }^2 } \right\rangle
			\approx
			\frac{ \left\langle\, \Gamma_{\mathrm{CR}} \,\right\rangle } { {\left\langle\, N_{\mathrm{CR}} \,\right\rangle}^2 }
			\qquad \forall \,\, \tau \ge \tau_{\mathrm{cc}}, 
			\label{eq:replenishment_condition}
	\end{equation}
	where the angle brackets denote long-term averages over time $t$, 
	$N_{\text{G}}(t;\,\tau)$ is the number of particles in the ghost system satisfying the same condition used to define $\tilde{\Gamma}(t;\,\tau)$, i.e., having no collision history in the interval $(t-\tau,\,t)$, while $N_{\text{CR}}$ ($N_{\text{CN}}$) is the number of particles in the CR (CN) system. In the middle segment of the above equation, the kernel is evaluated using the less conventional form $ \left\langle { \Gamma } /{ {N}^2 } \right\rangle$, as necessitated by the statistical non-stationarity of the CN system in which both $\Gamma$ and $N$ decay slowly with time; for the other two stationary systems, either definition yields essentially identical results, since $ \frac{ \left\langle\, \Gamma_{\mathrm{}} \,\right\rangle } { {\left\langle\, N_{\mathrm{}} \,\right\rangle}^2 } 
	\approx  \left\langle \frac{ \Gamma } { {N}^2 } \right\rangle $ in these cases (this is strongly corroborated by our data, see Appendix~\ref{app:ave_kernel}).
	The existence of $\tau_{cc}$ in this manner is supported by the findings presented earlier, since we observed in Fig.~\ref{fig:n2} that 
	$\frac{\tilde{\Gamma}_{\text{G}}(t;\, \tau=0)}{{\tilde{N}_{\text{G}}(t;\, \tau=0)}^2} 
	\equiv \frac{\Gamma_{\text{G}}(t) }{{N}_{\text{G}}(t)\,^2} 
	>   \frac{\Gamma_{\text{CN}}(t) } {N_{\text{CN}}(t)\,^2}  \approx  \frac{\Gamma_{\text{CR}}(t) }{N_{\text{CR}}(t)\,^2}$ for all $t >0$; 
	and conversely, Fig.~\ref{fig:n6} shows that in the opposing limit,
	$\frac{ \tilde{\Gamma}_{\text{G}}(t;\, \tau=\infty) } { {\tilde{N}_{\text{G}}(t;\, \tau=\infty)}^2 }  
	\approx 
	\frac{ \Gamma_{\text{CN}}(t) }{ N_{\text{CN}}(t)\,^2} 
	\approx  
	\frac{\Gamma_{\text{CR}}(t) }{N_{\text{CR}}(t)\,^2}$ for all $t$. 
	If collision enhancement due to correlation with history persists over a finite timescale, we expect that as $\tau$ increases from zero, $\left\langle \tilde{\Gamma}_{\text{G}}(\tau) \right\rangle$ will decay more rapidly than $ { \left\langle \tilde{N}_{\text{G}}(\tau) \right\rangle}^2$, bringing $\frac{\left\langle \tilde{\Gamma}_{\text{G}}(\tau) \right\rangle } { {\left\langle \tilde{N}_{\text{G}}(\tau) \right\rangle}^2}$ closer to $\left\langle\frac{ \Gamma_{\mathrm{CN}}  } { { N_{\mathrm{CN}} }^2 } \right\rangle$ and reaching it at a finite $\tau$. 
	
	Heuristically, the role of $\tau_{\text{cc}}$ may be illuminated through a thought experiment:  imagine a simulation where the colliding particles are masked (cloaked) for a period of $\tau$. While masked, they are advected like in a ghost-particle system but they do not participate in any collision until they are reinstated. When a sufficiently large $\tau$ is chosen, the reinstated particles are essentially decorrelated from their collision history and the reinstatement process itself mimics the spatially-random injection of new particles in the CR simulation. In view of this  and the preceding paragraph,  $\tau_{\text{cc}}$ may be interpreted as a collision-correlation time. 
	
	There is, however, another timescale of possible relevance which we shall call the injection-thermalization time. This timescale is defined with respect to the newly injected particles in the CR system. Since these particles are injected at purely random positions, their radial distribution function should be initially flat and equal to unity at all scales. In contrast, particles which have resided in the system for a long time are known to exhibit RDF with enhanced values at small scales. Thus, the injection-thermalization time may be understood as the characteristic timescale for the newly injected particles to develop correlations intrinsic to the particle-turbulence system. 
	A corollary of this is that the CR system always contains a fraction of "un-thermalized" particles, which results in $g_{\text{CR}}(r)$ being slightly leveled relative to $g_{\text{CN}}(r)$ and $\langle \frac{\Gamma_{\text{CR}}}{{N_{\text{CR}}}^2} \rangle$ being slightly smaller than $\langle \frac{\Gamma_{\text{CN}}}{{N_{\text{CN}} }^2} \rangle$, with the magnitude of these discrepancies increasing systematically with collision rate.
	The existence of this timescale suggests that it may also be possible to have the history-filtered coalescence rate mimic the coalescence rate in the CR system (i.e.,  $\tilde{\Gamma}_{\text{G}}(\tau)  \approx  \Gamma_{\text{CR}} $) for a specifically chosen $\tau$ that is close to the injection-thermalization time. The specific mechanics of this would however depend on the relation and relative magnitudes of the two timescales discussed (we defer this for future works). 
		 
\begin{figure}[htbp]
			\centering
			\includegraphics[width=0.55\linewidth]{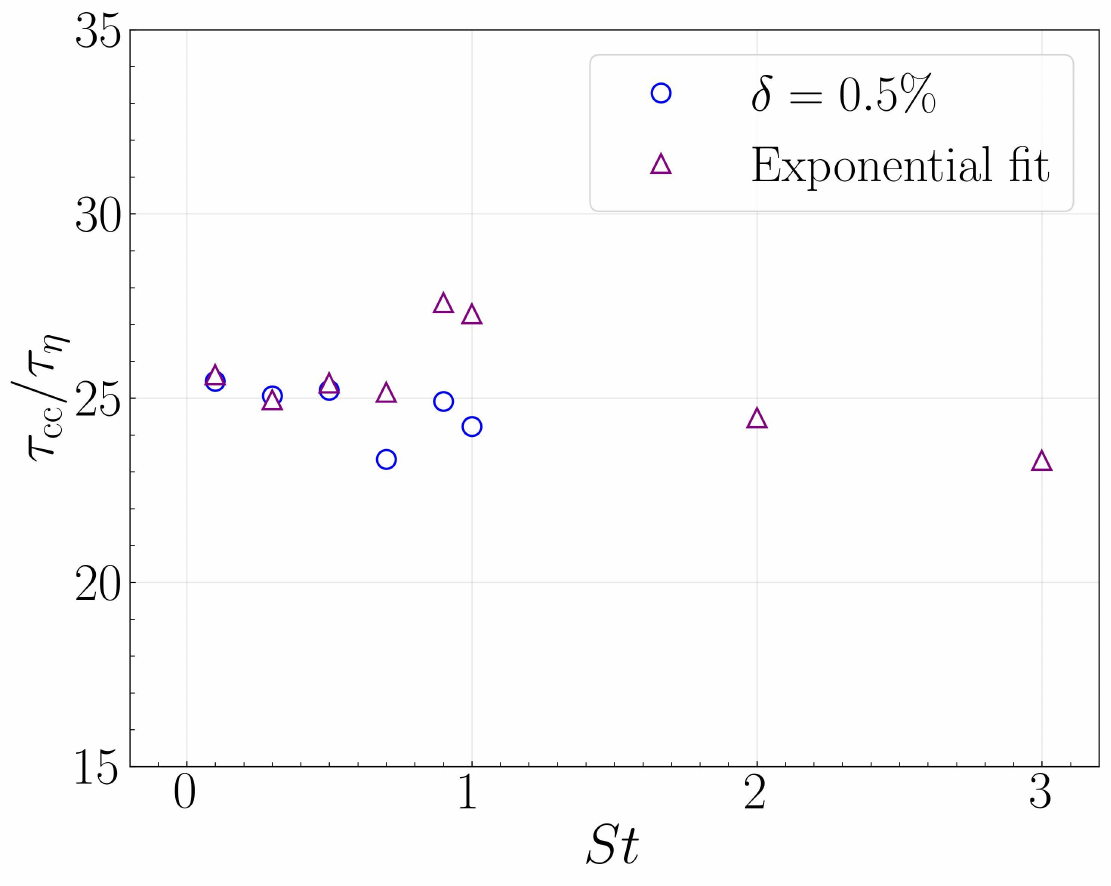}
			\caption{
				Collision-correlation time $\tau_{\mathrm{cc}}$ normalized by the Kolmogorov time scale $\tau_\eta$ as a function of the Stokes number ($St$); obtained from the threshold method with tolerance $\f{\delta} = 0.5\%$ and from exponential fits (details in text).
			}
			\label{fig:n7}
	\end{figure}

	We now turn to the magnitude of $\tau_{\text{cc}}$ (collision-correlation time) and its dependence on $St$.
	As shown in Fig.~\ref{fig:n7}, $\tau_{\mathrm{cc}}$ is of order $\tau_L$ (of order $10\tau_\eta$), indicating that the correlation between successive collisions is affected by large-scale turbulent eddies. Two different procedures were attempted to estimate $\tau_{\mathrm{cc}}$: (i) a threshold-based criterion, 
	where $\tau_{\mathrm{cc}}$ is identified as the smallest $\tau$ for which the percentage difference between $ f(\tau) \triangleq \frac{\left\langle \tilde{\Gamma}_{\text{G}}(\tau) \right\rangle } { {\left\langle \tilde{N}_{\text{G}}(\tau) \right\rangle}^2}$ and $\left\langle\frac{ \Gamma_{\mathrm{CN}}  } { { N_{\mathrm{CN}} }^2 } \right\rangle$ falls below a threshold $\epsilon = 0.5\%$; and (ii) an exponential curve fitting of the form   {$A \exp(-3\tau / \tau_{\mathrm{cc}}) + B$} against $f(\tau)$, where $A, B$, and   {$\tau_{\mathrm{cc}}$} are the results of the fit. We find that the threshold-based estimates work well only for $St \le 1.0$; for larger $St$, this method turns out to be less desirable possibly due to  stronger sampling noise inherent in $f(\tau)$ and $\left\langle\frac{ \Gamma_{\mathrm{CN}} } { { N_{\mathrm{CN}} }^2 } \right\rangle$.  
	Across all cases,   {$\tau_{\mathrm{cc}}$} lies in the narrow range of $23$--$28\,\tau_{\eta}$ ($\sim 2.5\tau_L$) with a possible weak dependence on $St$. A weak $St$-dependence (decay) may be consistent with the existence of the ballistic limit at $St \to \infty$ where particles are insensitive to turbulence-induced correlations.
	
	\subsection{History-Filtered Radial Distribution Function}
	
	We next examine the implication of $\tau_{\text{cc}}$, and the related history filtering methodology on clustering statistics as quantified by the RDF. To this end, we define a history-conditioned radial distribution function of the ghost particle system, denoted by $\tilde g_{\text{G}}(r;\,\tau)$. At each instant $t$, we only consider particles that have not experienced any collision within the interval $(t-\tau, \,t)$, and the RDF is evaluated over this conditioned particle ensemble using the standard pair-separation analysis. Subsequently, the instantaneous RDFs are averaged over time (over the last $140 \tau_{\eta}$ of the simulation) for the sake of statistical convergence. 
			
		\begin{figure}[htbp]
				\centering
				\includegraphics[width=0.55\textwidth]{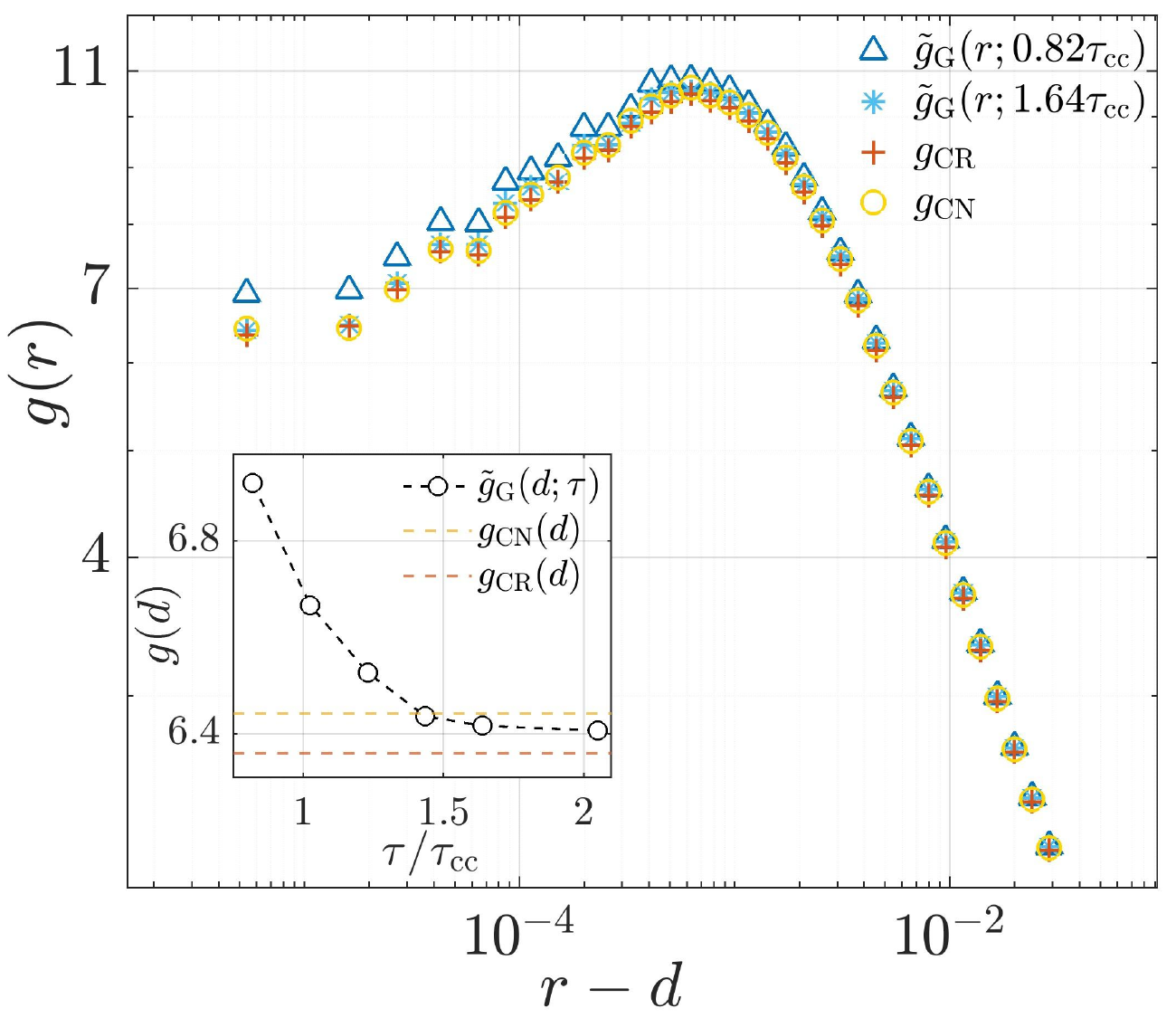}
				\caption{ 
				Effect of history-filtering on the radial distribution function for the representative case of $St=0.5$ and initial number of particles $N(t\,{=}\,0)=2.5\times10^6$. The main panel shows the RDFs $g(r)$ as a function of gap distance $r-d$, including the history-filtered results $\tilde g_{\text{G}}(r; \, 0.82\tau_{\text{cc}})$ and $\tilde g_{\text{G}}(r; \, 1.64\tau_{\mathrm{cc}})$ ($\tau_{\text{cc}}$ being the collision-correlation time), along with the coalescing systems RDFs $g_{\text{CR}}$ and $g_{\text{CN}}$ as benchmarks. We see that $\tilde g_{\text{G}}(r; \, \tau)$ can reproduce $g_{\text{CR}}$ and $g_{\text{CN}}$ for sufficiently large filter-window $\tau$ of the order of $\tau_{\text{cc}}$.  
				\textbf{Inset)} The RDF value at particle contact $\tilde g_{\text{G}}(d; \, \tau)$ as a function of the filter-window $\tau$, with horizontal dashed lines indicating $g_{\text{CR}}(d)$ and $g_{\text{CN}}(d)$. This shows how the history-filtered RDF approaches the benchmarks as $\tau$ is increased. 
				}
				\label{fig:n8_tcc}
		\end{figure}
		
		\begin{figure}[htbp]
			\centering
			\includegraphics[width=1\textwidth]{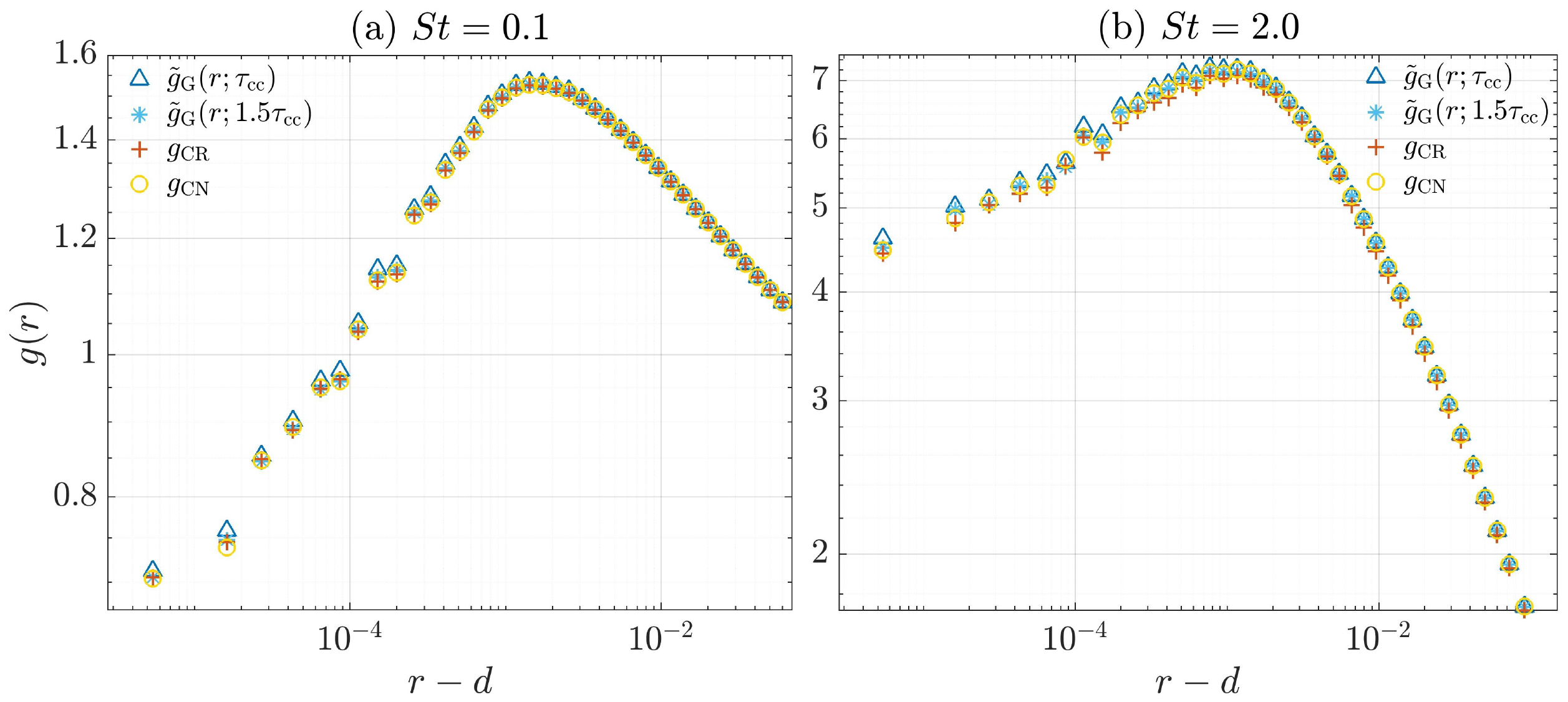}
			\caption{ 
			History-filtered RDFs for additional Stokes numbers: (a) $St=0.1$ and (b) $St=2.0$. The history-filtered ghost-particle RDFs are compared with the corresponding coalescing-system RDFs.
			}
			\label{fig:n9_tcc}
		\end{figure}
		The resulting history-filtered RDF, $\tilde{g}_{\text{G}}(r;\,\tau)$, for the representative case of $St = 0.5$, is shown in Figure~\ref{fig:n8_tcc}. 
		Two different values of $\tau$ of order $\tau_{\mathrm{cc}}$ were attempted and the outcomes were compared with the collision--coalescence benchmarks $g_{\text{CR}}$ and $g_{\text{CN}}$. For $\tau=0.82\tau_{\mathrm{cc}}$, the conditioned RDF 
		is already close to the CN/CR curves, although a residual excess of roughly $7.5\%$ remains at near-contact separations  ($r \approx d$). Increasing $\tau$ to $1.64\tau_{\mathrm{cc}}$ further reduces the discrepancy, making the history-filtered RDF nearly indistinguishable from $g_{\text{CR}}$ and $g_{\text{CN}}$. 
		To assess the generality of the proposed history-filtering procedure, Figure~\ref{fig:n9_tcc} presents analogous results for $St=0.1$ and $St=2.0$.
		In both cases, increasing the filter window to $1.5\tau_{\mathrm{cc}}$	brings the history-filtered RDF progressively closer to the corresponding coalescing-system RDFs.
		Here, it is worth noting that $g_{\text{CN}}(r)$ is, to leading order, equal to $\tilde g_{\text{G}}(r;\,\tau \to \infty)$. Thus, we expect that $\tilde{g}_{\text{G}}(r;\,\tau)$ does not change significantly as $\tau$ is increased to larger values. Indeed, the behavior of $\tilde{g}_{\text{G}}(r;\,\tau)$ at larger $\tau$ shown in Appendix~\ref{app:rdf_taucc}  strongly corroborates this expectation. 
	
		The inset in Fig.~\ref{fig:n8_tcc} shows how the magnitude of RDF at contact, $\tilde g_{\text{G}}(d;\,\tau)$, evolves as a function of $\tau$. 
		It shows that as $\tau$ increases, $\tilde g_{\text{G}}(d;\,\tau)$ decreases rapidly at first but later plateaus to a value close to $g_{\text{CR}}(d)$ and $g_{\text{CN}}(d)$, reinforcing the idea that $\tau_{\text{cc}}$ is a threshold beyond which the history-filtered statistics of the ghost-particle system (in this case the RDF) reproduces the corresponding statistics in the coalescing systems. The above observations demonstrate the feasibility of a theoretical bridge between the two system types. 
	
		Figure~\ref{fig:n8_tcc}, arguably the centerpiece result of this work, together with the additional demonstrations in Fig.~\ref{fig:n9_tcc}, brings
		us full circle back to the challenge inspired by Fig.~\ref{fig_teaser}, that is, to understand the difference between the ghost and coalescing systems' RDFs and to reproduce the latter using an altered version of the former. This result strongly suggests that the difference is the consequence of temporal correlation among successive collisions experienced by a single particle and the figure clearly shows that the history-filtered RDF $\tilde g_{\text{G}}(\tau \gtrsim \tau_{\text{cc}})$ could reproduce $g_{\text{CR}}$ and $g_{\text{CN}}$. In retrospect, the velocity-filtered RDF $g_{\text{G}}^{\scriptscriptstyle (-)}(r)$ featured in Fig.~\ref{fig_teaser} may be viewed as simply a specialized version of the history-filtered RDF $\tilde g_{\text{G}}(r;\,\tau)$ with a small $\tau$ (of the order of transit time for ghost-particle overlap). 
		
		The above findings provide strong evidence for the elegant conjecture that all statistical discrepancies between coalescing and ghost particle systems arise solely from collision correlation or collision memory. In this picture, a coalescing system may be seen as a ghost-particle system with short-term memory lost\footnote{A more complete account of this statement can be found in the thought experiment discussed earlier.}. This view unifies the two systems under a common framework and paves the way for prediction of coalescing system statistics from the ghost-particle framework.
		The prediction of coalescing particle statistics will be a subject for our future works. 
	
	\subsection{Partner Persistency in Correlated Collisions}
		The previous analysis demonstrates the significance of correlation between successive collision events without the need to consider whether two correlated collision events involve the same or different pairs of particles. As a point of interest, here we examine the persistence of particle pairing by calculating the fraction of re-collisions involving identical particles among all collision events that have collision history  $\langle \Gamma_{\mathrm{\scriptscriptstyle G}}^{ hs} / \Gamma_{\mathrm{\scriptscriptstyle G}}^{ h}  \rangle$, where $\Gamma_{\mathrm{\scriptscriptstyle G}}^{ h} = \Gamma_{\text{G}} - \tilde{\Gamma}_{\text{G}}$ denotes the rate of collision conditioned on at least one of the colliding particles having a collision history, and $\Gamma_{\mathrm{\scriptscriptstyle G}}^{ hs}$ denotes the rate of collision conditioned on the particles having previously collided with each other in their most recent collision. We note that collisions involving more than two particles are extremely rare under current density and time resolution, and the phenomenon is beyond the scope of the current work. 
		
		\begin{figure}[htbp]
			\centering
			\includegraphics[width=0.55\linewidth]{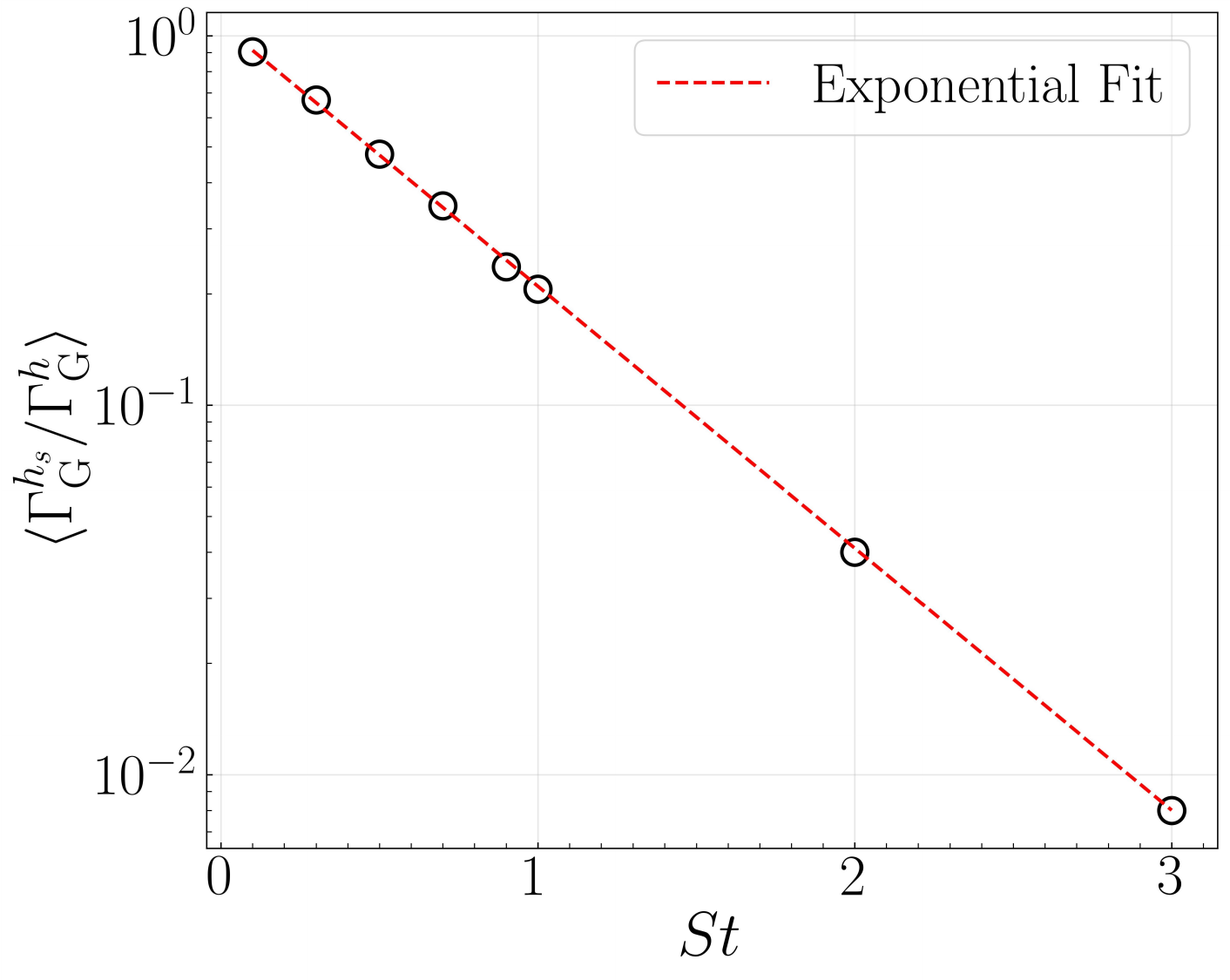}
			\caption{
				Stokes number dependence of the fraction of identical pair re-collision events among history-bearing collision events. Black open circles denote the DNS outcomes for $\langle \Gamma_{\mathrm{\scriptscriptstyle G}}^{ hs} / \Gamma_{\mathrm{\scriptscriptstyle G}}^{ h} \rangle$, at various $St$. The red dashed line shows an exponential fit, $Y = 1.075 \, \exp(-1.633 \, X)$.
		}
		\label{fig:n10}
		\end{figure}
		Figure~\ref{fig:n10} presents the Stokes-number dependence of $\langle \Gamma_{\mathrm{\scriptscriptstyle G}}^{ hs} / \Gamma_{\mathrm{\scriptscriptstyle G}}^{ h}  \rangle$. We see that this fraction decreases approximately exponentially with $St$, implying that for less inertial particles, repeated encounters are more frequent. 
		This suggests the tendency of these particles to remain temporarily paired due to persistent spatial clustering and coherent trapping by small-scale turbulent structures. As particle inertia increases, 
	    successive collisions become more well mixed with respect to partnering.
		
	\section{CONCLUSION}\label{sec:con}
	
		This study provides a unified interpretation of the discrepancy of collision and clustering statistics between the systems of coalescing inertial particles and inertial ghost-particle in turbulence, and shows that it arises from correlation among successive collisions inherent to the ghost-particle framework.
		
		Using three types of direct numerical simulations, namely those involving ghost particles (GP), colliding-coalescing particles with replenishment of lost monomers (CR), and colliding-coalescing particles without monomer replenishment (CN), we showed that both the collision kernel $K$ and the radial distribution function near contact, $g(r \approx d)$, of the coalescing systems are consistently lower than the GP system across the range of Stokes numbers considered ($St = 0.01$--$3.0$).
		We found that the velocity-filtered RDF, $g_{\mathrm{G}}^{(-)}(r)$, serves as a reasonable but imperfect proxy for $g_{\mathrm{CR}}(r)$ at near contact particle separations. We systematically quantify the residual discrepancy in RDF and collision kernel between the velocity-filtered ghost-particle and coalescing particle results. 
		We define a novel history-filtered collision kernel for the GP system, $\tilde{\Gamma}_{\mathrm{G}}$, and show that it reproduces the collision kernel of the coalescing systems, thus strongly suggesting that the aforementioned discrepancy is due to correlation among successive collisions in the GP system.
	
		We generalize the concept of history-filtering and define the parametrized history-filtered kernel $\tilde{\Gamma}_{\mathrm{G}}(t;\, \tau)$. This led to the identification of the collision-correlation timescale,   {$\tau_{\text{cc}}$}, which quantifies the temporal extent over which successive collisions remain correlated. We find that   {$\tau_{\text{cc}}$} has a narrow range in the tenths of Kolmogorov timescale of the flow and exhibits a weak dependence on Stokes number. Similarly, we define a novel history-filtered RDF $\tilde{g}_{\mathrm{G}}(r;\,\tau)$. As the filtering window $\tau$ increases, both the history-filtered kernel and RDF of the ghost-particle system progressively approach those of the coalescing systems and eventually saturate at values close to the latter once correlation effects are removed. We argued that the velocity-filtered $g_{\mathrm{G}}^{(-)}(r)$ is a specialized member of the more general set of $\tilde{g}_{\mathrm{G}}(r;\,\tau)$. 
	
		Separately, we found that the fraction of repeated collisions between identical particle pairs decreases exponentially with increasing Stokes number, indicating a diminishing role of pairwise re-collisions at higher inertia.
		The results in this work unify the coalescing and ghost particle systems under a single framework by providing a theoretical bridge between them. Specifically, they afford an elegant interpretation of the coalescing particle system as a history or memory filtered version of the ghost-particle system. This paves the way for the prediction of coalescing system statistics from the GP framework. 
		
	\section{ACKNOWLEDGEMENTS}
		We thank Xiaohui Meng for various discussions on this work. FG is especially grateful for her support and encouragement. We thank J\'er\'emie Bec for insightful discussions. We also thank Jun Feng, Linli Fu, Yating Chen, Li Zhang for numerous discussions and support. This work was mainly supported by the National Natural Science Foundation of China (Grant No. 11872382) and by the Thousand Young Talents Program of China.
		
	\setcounter{figure}{0}
	\renewcommand{\thefigure}{A\arabic{figure}}
	\setcounter{table}{0}
	\renewcommand{\thetable}{A\arabic{table}}
	
	\appendix
	\section{Particle Replenishment}
	\label{app:injfreq}
\begin{figure}[htbp]
		\centering
		\includegraphics[width=0.75\linewidth]{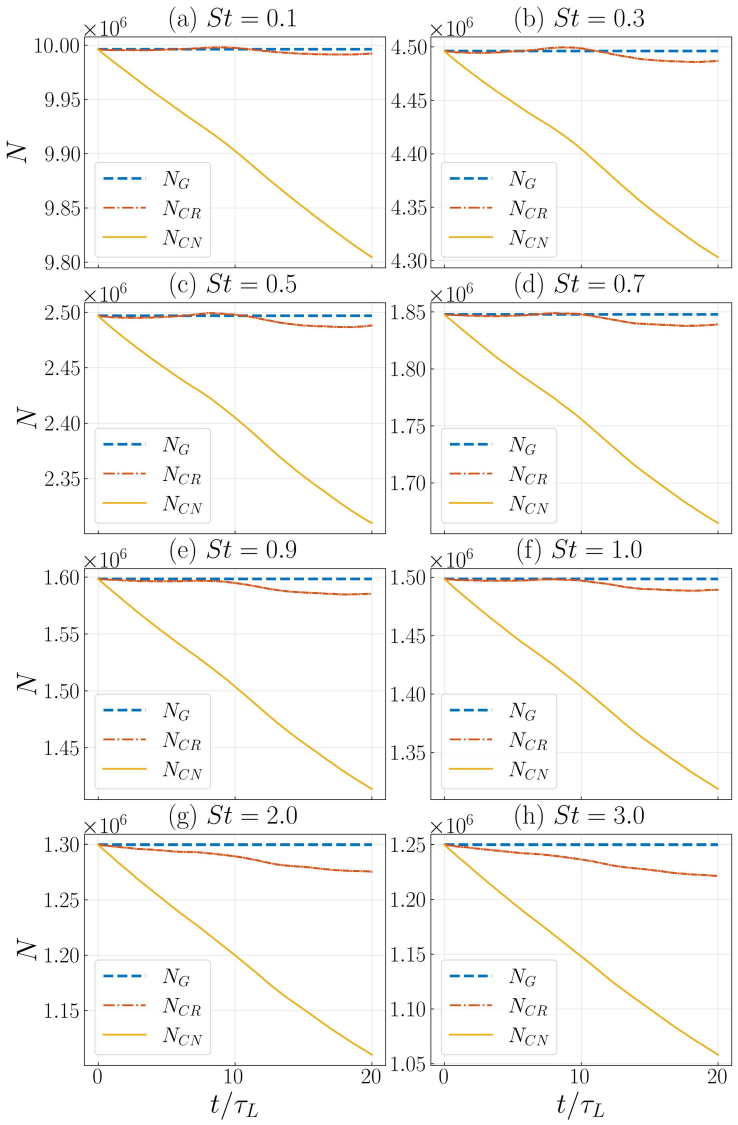}
		\caption{
		Time evolution of the relative deviation of particle number in CR and GP system cases for different Stokes numbers.}
		\label{fig:A1}
	\end{figure}

	In the CR case, new particles were injected periodically at random positions uniformly over the entire computational domain, so as to maintain a nearly constant particle number. Across all Stokes numbers, particles were injected at a fixed rate of 96 particles every $0.1\tau_\eta$. The instantaneous particle number $N$ was continuously monitored to evaluate the effectiveness of the replenishment protocol \g{(see Fig.~\ref{fig:A1})}.
	
	The relative deviation of the instantaneous particle number between the CR and GP system cases was quantified as $|N_{\mathrm{CR}} - N_{\mathrm{G}}| / N_{\mathrm{G}}$.
	After applying a uniform injection protocol across all Stokes numbers, the deviation remains below 0.01 for $St \le 1$. For higher-inertia particles, the deviation increases gradually over time,
	reaching a maximum of approximately 0.022 for $St = 3$ after 20$\tau_L$.
	This small accumulated deviation suggests that the injection scheme maintains a particle population sufficiently close to the ghost-particle value, without significantly affecting the collision statistics.
	
	\section{Effect of order of operations on kernel evaluation}
	\label{app:ave_kernel}
	For the two statistical stationary systems, we compare the two formulae $\langle \Gamma/N^2 \rangle$ and $\langle \Gamma \rangle / \langle N \rangle^2$ for computing the reduced kernel $\Gamma/N^2$.
	The relative difference is evaluated as the absolute difference between the two quantities, normalized by $\langle \Gamma \rangle / \langle N \rangle^2$ and expressed in percentage.
	Table~\ref{tab:kernel_av} summarizes representative results for $\Gamma_{\mathrm{CR}} / {N_{\mathrm{CR}}}^2$ and $\tilde{\Gamma}_{\mathrm{G}}(t;\tau) / [\tilde{N}_{\mathrm{G}}(t;\tau)]^2$ with $\tau = 30\tau_\eta$. For $\Gamma_{\mathrm{CR}} / {N_{\mathrm{CR}}}^2$, the relative difference remains below $0.019\%$ across all $St$, while for $\tilde{\Gamma}_{\mathrm{G}}(t;\tau) / [\tilde{N}_{\mathrm{G}}(t;\tau)]^2$ it is below $0.002\%$. This confirms that the difference between the two definitions is negligible for both stationary systems.
	Therefore, the two formulations are numerically indistinguishable for the present simulations.
	\begin{table}[h]
		\caption{Representative comparison of $\Gamma_{\mathrm{CR}} / {N_{\mathrm{CR}}}^2$ and $\tilde{\Gamma}_{\mathrm{G}}(t;\tau) / [\tilde{N}_{\mathrm{G}}(t;\tau)]^2$ with $\tau = 30\tau_\eta$, comparison between $\langle \Gamma/N^2 \rangle$ and $\langle \Gamma \rangle / \langle N \rangle^2$. All values are scaled by \fs{$10^{-11}$}.}
		\label{tab:kernel_av}
		\centering
		\begin{ruledtabular}
		\renewcommand{\arraystretch}{1.5}
		\begin{tabular}{ccccc}
			$St$ & $\langle \Gamma_{\mathrm{CR}} / {N_{\mathrm{CR}}}^2 \rangle$ & $\langle \Gamma_{\mathrm{CR}} \rangle / \langle N_{\mathrm{CR}} \rangle^2$ & $\langle \tilde{\Gamma}_{\mathrm{G}} / [\tilde{N}_{\mathrm{G}}]^2 \rangle$ & $\langle \tilde{\Gamma}_{\mathrm{G}} \rangle / \langle \tilde{N}_{\mathrm{G}} \rangle^2$ \\
			\hline
			0.1 & 4.948 & 4.949 & 4.9528 & 4.9527 \\
			0.5 & 82.219 & 82.218 & 82.868 & 82.849 \\
			1.0 & 228.502 & 228.503 & 231.077 & 231.038 \\
			3.0 & 365.24 & 365.27 & 365.53 & 365.51 \\
		\end{tabular}
		\end{ruledtabular}
		\end{table}
		
	\section{History-filtered RDF at Larger Filter Windows}
	\label{app:rdf_taucc}
	To assess the dependence on the history window $\tau$, we show, in Fig.~\ref{fig:A2}, the history-filtered RDF $\tilde{g}_{\mathrm{G}}(r;\,\tau)$ of the $St=0.5$ case for $\tau = 1.64\,\tau_{\mathrm{cc}}$, $2.05\,\tau_{\mathrm{cc}}$, and $5\,\tau_{\mathrm{cc}}$. 
	The three curves collapse onto each other, indicating saturation with respect to $\tau$. {To further verify the robustness of this trend, we examined the history-filtered RDF for $St = 0.1$ and $St = 2.0$ across filter windows $\tau = 1.5\,\tau_{\text{cc}}$, $2\,\tau_{\text{cc}}$, and $5\,\tau_{\text{cc}}$. As shown in Fig.~\ref{fig:A3}, the curves for different $\tau$ values exhibit an almost perfect collapse. These results indicates that the RDFs reach saturated states at $\tau \approx   {\tau_{\text{cc}}}$ and is insensitive to further increases in $\tau$ across a range of particle Stokes number.
	
	\begin{figure}[htbp]
		\centering
		\includegraphics[width=0.55\linewidth]{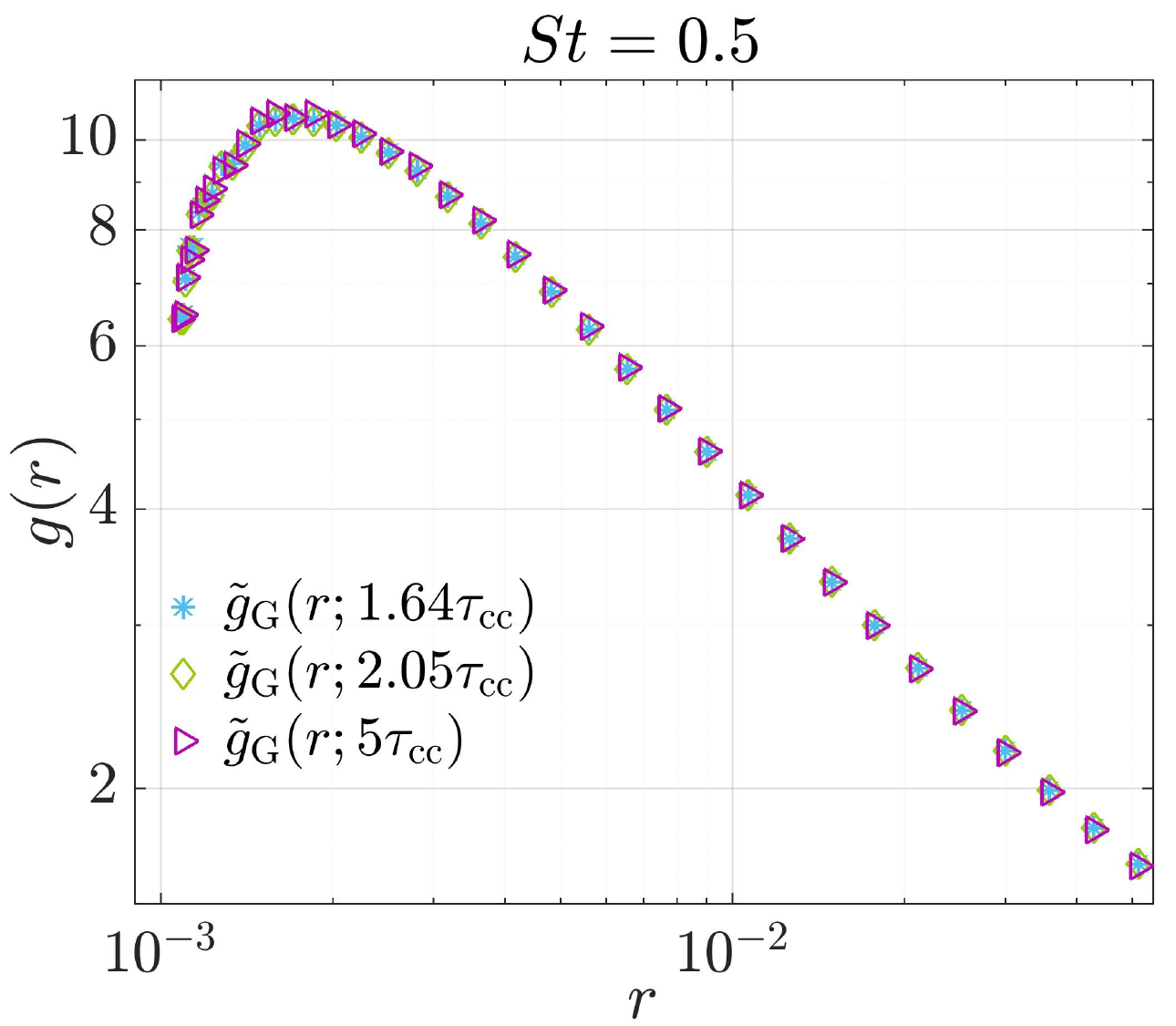}
		\caption{
			History-conditioned RDF $\tilde{g}_{\mathrm{G}}(r;\,\tau)$ at various $\tau$ values for $St=0.5$.
		}
		\label{fig:A2}
	\end{figure}
	
	\begin{figure}[htbp]
		\centering
		\includegraphics[width=1\linewidth]{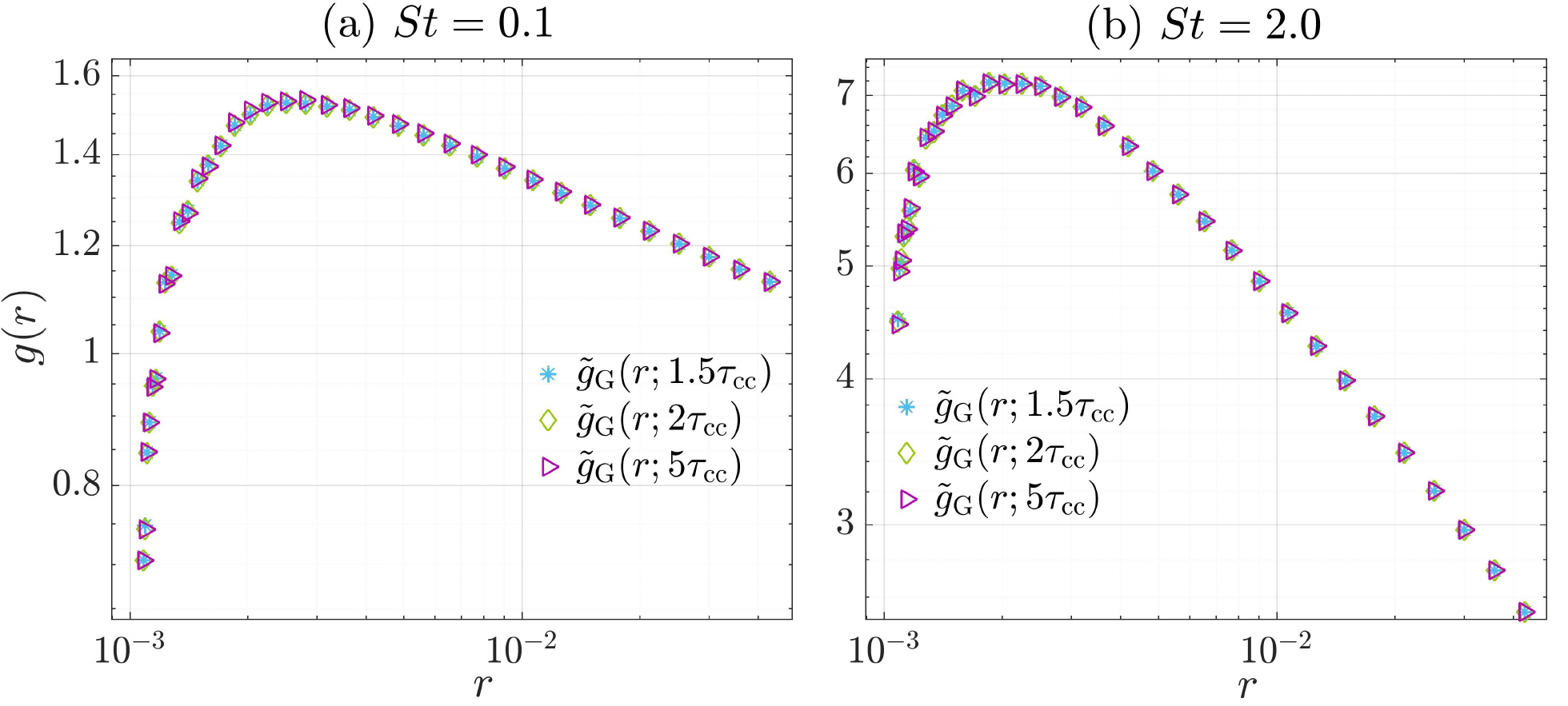}
		\caption{
			History-conditioned RDF $\tilde{g}_{\mathrm{G}}(r;\,\tau)$ at various $\tau$ values: (a) $St=0.1$, (b) $St=2.0$.
		}
		\label{fig:A3}
	\end{figure}
\newpage
\bibliography{ghref.bib}


\end{document}